\documentclass[12pt, draftcls, onecolumn]{IEEEtran}
\usepackage{graphicx}
\usepackage{amsmath,amssymb}
\usepackage[noadjust]{cite}
\usepackage{psfrag}
\usepackage{graphics}
\usepackage{graphicx}
\usepackage{times}
\usepackage{color}
\usepackage{soul}
\newfont{\bboard}{msbm10 scaled\magstephalf}

\newtheorem{theorem}{Theorem}[section]

\newtheorem{corollary}[theorem]{Corollary}
\newtheorem{definition}[theorem]{Definition}
\newtheorem{lemma}[theorem]{Lemma}
\newtheorem{proposition}[theorem]{Proposition}

\begin{document}
\title{A Passivity Framework for Modeling and Mitigating Wormhole Attacks on Networked Control Systems}
\author{Phillip Lee, Andrew Clark, Linda Bushnell\footnote{$^{1}$Corresponding author.  Email: lb2@uw.edu}$^{1}$, and Radha Poovendran \\
Dept. of Electrical Engineering, University of Washington, Seattle, WA, 98195, USA \\
\{leep3, awclark, lb2, rp3\}@uw.edu}
\maketitle
\begin{abstract}
Networked control systems consist of distributed sensors and actuators that communicate via a wireless network.  The use of an open wireless medium and unattended deployment leaves these systems vulnerable to intelligent adversaries whose goal is to disrupt the system performance.  In this paper, we study the wormhole attack on a networked control system, in which an adversary establishes a  link between two distant regions of the network by using either high-gain antennas, as in the out-of-band wormhole, or colluding network nodes as in the in-band wormhole. Wormholes allow the adversary to violate the timing constraints of real-time control systems by delaying or dropping packets, and cannot be detected using cryptographic mechanisms alone.  We study the impact of the wormhole attack on the network flows and delays and introduce a passivity-based control-theoretic framework for modeling the wormhole attack.    We develop this framework for both the in-band and out-of-band wormhole attacks as well as complex, hereto-unreported wormhole attacks consisting of arbitrary combinations of in-and out-of band wormholes. We integrate existing mitigation strategies into our framework, and analyze the throughput, delay, and stability properties of the overall system.  Through simulation study, we show that, by selectively dropping control packets, the wormhole attack can cause disturbances in the physical plant of a networked control system, and demonstrate that appropriate selection of detection parameters mitigates the disturbances due to the wormhole while satisfying the delay constraints of the physical system. 


\end{abstract}

\section{Introduction}
\label{sec:intro}
Cyber-physical systems that are deployed over a wide geographic area often consist of distributed embedded devices, such as sensors and actuators, that exchange sensed data and control signals via a wireless network~\cite{pajic2011wireless}, thus forming a networked control system.  When deployed in critical applications such as the smart grid, the real-time control system may be targeted by adversaries attempting to drive it to an undesirable or unsafe operating point. By introducing and modifying delays in the communication network, the adversary can cause violations of the timing constraints that are critical in maintaining safe operation of real-time cyber-physical systems \cite{jahanian1986safety}.

The wormhole attack, first introduced in the context of wireless routing \cite{hu2003packet}, is one such attack that exploits the time delays and violates the timing constraints of the targeted system.
In the wormhole attack, an adversary records messages observed in one region of the network and replays them in a different region~\cite{karlof2003secure}.  By doing so, the adversary creates a communication link (a wormhole tunnel) between two end points in  otherwise disjoint geographic areas. This can be accomplished by either compromised or colluding network nodes, known as the in-band wormhole \cite{kruus2006band} or via a side channel such as high-gain directional antennas, known as the out-of-band wormhole \cite{hu2003packet}.  Unsuspecting network nodes will route network traffic through the wormhole. Once significant traffic starts flowing through the wormhole, the adversary can selectively drop or delay time-critical packets in order to destabilize or degrade the system performance. As the attack replays or reroutes valid messages, it does not require compromising any cryptographic keys, and hence cannot be detected using cryptographic verification mechanisms alone~\cite{poovendran2007graph}.

While the wormhole attack does not violate cryptographic mechanisms, it does violate the physical constraints imposed by propagation delay and relative position of nodes.  Current approaches that detect these violations include include graph-based methods \cite{poovendran2007graph}, statistical methods \cite{kruus2006band}, and timing analysis \cite{hu2003packet}. However, the current security analysis of the mitigation strategies do not incorporate the time-varying node behaviors or the adaptive strategy of the adversary. Hence, while the wormhole attack can significantly degrade the performance of cyber-physical systems, there is currently no analytical approach that represents the impact of wormholes and mitigation on the system dynamics.  Furthermore, the composition of different types of wormhole attacks and the impact on system performance has not been studied.



In this paper, we introduce one such control-theoretic framework for modeling and mitigating the wormhole attack on networked control systems. The proposed framework models the impact of wormholes, as well as the integration of existing mitigation strategies, on the allocation of network flows and resulting delays. Our approach models  three interdependent components, namely, flow allocation by network nodes, delay characteristics introduced by wormholes, and mitigation algorithms employed by the network. We develop this framework for both out-of-band and in-band wormholes.  In addition, using our framework, we  are able to model, represent  and mitigate  complex wormhole attacks that simultaneously make use of both in- and out-of-band wormholes. For each case, we prove that the flow allocation, wormhole delay, and mitigation components can be modeled as a passive dynamical system which allows the characterization of flow allocation and delay at the steady state.  Since our framework is in control-theoretic language, it enables ease of composition with control models of cyber-physical systems.
We make the following specific contributions:
\begin{itemize}
 \item For the out-of-band wormhole, we develop dynamical models for the flow allocation by network nodes, the delays introduced by wormholes, and network mitigation. For the flow allocation by network nodes, we introduce a distributed algorithm for each node to adaptively divide its flow among a set of paths based on their delays. We prove this algorithm converges to a unique Wardrop equilibrium, in which no source can reduce its delay by shifting flow to a different path.
 \item We model the delay characteristics of out-of-band wormhole links based on the packet dropping rate. We map the packet dropping strategy to the optimization problem of selecting the optimal dropping rate which balances the goals of increase in delay and attracting flows to the wormhole. We then develop a dynamical model to characterize the effect of timing based mitigation mechanisms on the flow allocation.
 \item We then prove the dynamical systems describing the flow allocation, delays introduced by wormhole, and the mitigation schemes are passive. We leverage the passivity property to prove that the interconnection of these models is globally asymptotically stable with respect to a unique equilibrium point.
 \item For the in-band wormhole, we derive the delays introduced by wormhole as a function of number of colluding nodes and the network topology. We represent statistics based mitigation method against in-band wormhole as a penalty added to suspected wormhole links during the flow allocation. We then prove that the flow allocation algorithm introduced earlier together with the delay and mitigation models in the in-band case, can be represented as an interconnection of passive systems, which converges to a stable equilibrium point.
 \item We use our framework to model more complex wormhole attacks, which consists of both in- and out-of-band wormholes.  Our approach composes the models of individual wormhole links via parallel interconnections of passive systems.
\item We illustrate our approach via a numerical study, in which we compare the flow allocation and delay resulting from both out-of-band and in-band wormhole attacks and the detection mechanisms, and evaluate the impact of the wormhole attack and mitigation on a cyber-physical system. In the out-of-band case, simulation results show that detection mechanisms reduces the flow traversing through the wormhole link at the cost of increased delay. For the in-band case, simulation results suggest that detection mechanisms enable the source rates to converge to the same equilibrium regardless of the presence of a wormhole.  We find that an adversary who creates an out-of-band wormhole can cause large disturbances on the physical plant by selectively dropping packets that are allocated to the wormhole link.  We empirically determine parameters of the mitigation strategy that reduces the flow allocated to the wormhole link, while satisfying the system's delay constraints.
\end{itemize}

Our proposed framework enables quantitative analysis of the impact of the wormhole attack on system performance and the effectiveness of different mitigation mechanisms, as well as modeling of any arbitrary composition of in-band and out-of-band wormholes.  Hence, this approach is complementary to recent efforts towards a science of cyber-security~\cite{schneider2011blueprint}, where the goal is a scientific approach to characterizing, composing, and mitigating security threats.  Moreover, our proposed framework explicitly captures the temporal dynamics of the attack and mitigation, including the adaptation and co-evolution of the adversary and defender strategies.

The paper is organized as follows.  We present the related work in  Section \ref{sec:related}.    Section \ref{sec:model} presents our assumptions of the network and adversary capabilities, as well as a description of the wormhole attack.  Section \ref{sec:approach} discusses our proposed modeling and mitigation framework for the out-of-band wormhole.  Section \ref{sec:in_band} presents our approach to modeling and mitigating in-band wormholes.  Section \ref{sec:joint} introduces passivity-based models and mitigation for joint out-of- and in-band wormholes.  Numerical results are contained in Section \ref{sec:simulation}. Section \ref{sec:conclusion} concludes the paper.  Appendix \ref{sec:passive_background} presents background on passivity. To improve readability, some of the lengthier proofs of our results are contained in  Appendix \ref{sec:proofs}.

\section{Related Work}
\label{sec:related}
The wormhole attack was originally identified as a form of routing misbehavior in ad hoc and sensor networks~\cite{karlof2003secure}.  In \cite{hu2003packet}, the packet leash defense was proposed, in which each packet is given a fixed expiration time and any packet received after its expiration time is discarded.  Valid packets may also be discarded, however, due to propagation delays or clock skews between nodes, leading to a trade-off between detection effectiveness and network performance.  Local broadcast keys, which are cryptographic keys that are distributed using specialized guard nodes and known only to nodes within a local neighborhood, were introduced in \cite{poovendran2007graph}.  Anomalies in link delays, caused by propagation through the wormhole tunnel, are analyzed in \cite{song2011enhancement}, in which an FFT-based approach to identifying likely wormholes was presented. While these methods can be used to mitigate the impact of the wormhole attack, an analytical approach to dynamically tune each method in response to changes in the network state and adversary behavior, as well as estimate the stable operating point of the system, is currently lacking.

The in-band wormhole, in which the adversary creates the appearance of a link between two colluding nodes by tunneling packets through valid nodes, was identified as a security threat in \cite{kruus2006band}.  The authors observed that the wormhole tunnel itself could contain routing loops, diminishing its effectiveness, a phenomenon they denoted as wormhole collapse.
  Necessary and sufficient conditions for the adversary to avoid wormhole collapse are derived in \cite{mahajan2008analysis}.  A statistical approach to detecting in-band wormholes, based on identifying increased delays or packet drops through wormhole links using sequential probability ratio testing, was studied in \cite{baras2007intrusion}.  Our framework incorporates the probability of wormhole collapse, as well as the statistical detection algorithms, when modeling the temporal dynamics of the flow rates and resulting delays.

 Passivity-based techniques have been used to model network flow control and derive novel flow allocation algorithms in \cite{wen2004unifying}.  The work of \cite{wen2004unifying} fits within the broader context of dual decomposition-based methods for designing network protocols as distributed algorithms for solving network optimization problems \cite{chiang2007layering}. Passivity of networked control systems with packet drops was studied in \cite{wang2012passivity}. In \cite{wang2012passivity}, the authors studied the passivity of networked control systems where the plant dynamics switches between open and closed loop due to control packet drops. Currently, however, such models do not incorporate security threats or network defenses.


 In preliminary versions of this work~\cite{clark2012passivity,lee2013modeling}, we studied passive dynamical systems as a framework for modeling and mitigating network security threats. In \cite{clark2012passivity}, we presented passive dynamical models of the node capture, malware propagation, and control channel jamming attacks, and demonstrated that these attacks can be composed while preserving passivity.  In \cite{lee2013modeling}, we studied a class of adaptive network defense mechanisms against control channel jamming that satisfy the passivity property, and demonstrated that the robustness of the system to delays and detection errors is affected by the parameters of the passive defense.  Neither of these works, however, consider network flow-based attacks such as the wormhole attack.

\section{Preliminaries}
\label{sec:model}
In this section, we state our assumptions regarding the capabilities of the network and adversary.  We then give background on the wormhole attack.
\subsection{Network Model}
We consider a wireless network of $n$ nodes.
We assume that two nodes can communicate directly if their positions are within the maximum node communication range.  We denote the set of links by $\mathcal{L}$, with $|\mathcal{L}| = L$. In order to facilitate sensing and control of the system, network flows must be maintained between a set of source nodes  $\mathcal{S}$ and destination nodes $\mathcal{D}$. The ordered pair $(S_{i}, D_{i})$ denotes the source and destination of flow $i$.  We assume that source $S_{i}$ maintains a constant rate $r_{i}$, and that flows are originating from the set of sources. External flows are not considered in this paper.

Any source and destination pair that is not in direct radio range relies on multi-hop communication. Since the topology changes due to node sleep/wake cycles and nodes joining and leaving the network, each source $S_{i}$ uses a distributed routing protocol to identify a set of source-destination paths $\mathcal{P}_{i} = \{P_{1}, \ldots, P_{m_{i}}\}$. The number of paths for source-destination pair ($S_i$, $D_{i})$ is denoted as $m_i$.

\subsection{Adversary Model}
The network is deployed in a hostile environment where one or more mobile adversaries are present.
We assume that each adversary  is capable of eavesdropping as well as recording and replaying eavesdropped messages, including routing protocol messages.  By eavesdropping on routing protocol messages, the adversary determines the network topology. The adversary is also capable of physically capturing the unattended nodes.
Once the adversary has compromised a node, the adversary can extract its cryptographic secrets. This enables the adversary to replace the captured node with a malicious node assuming the identity of the captured node.
Malicious nodes are under the control of the adversary and are capable of colluding with other malicious nodes.
One such collusion attack is the wormhole, described as follows.
\begin{figure*}[h]
\centering
$\begin{array}{cc}
\includegraphics[width=3in]{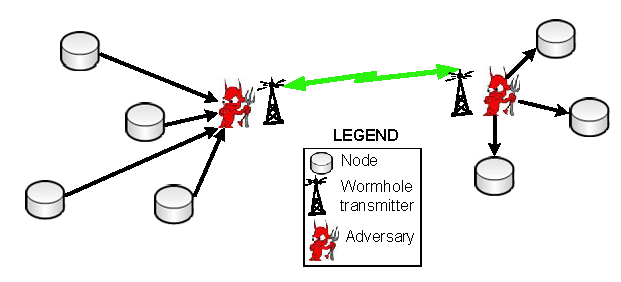} &
\includegraphics[width=3in]{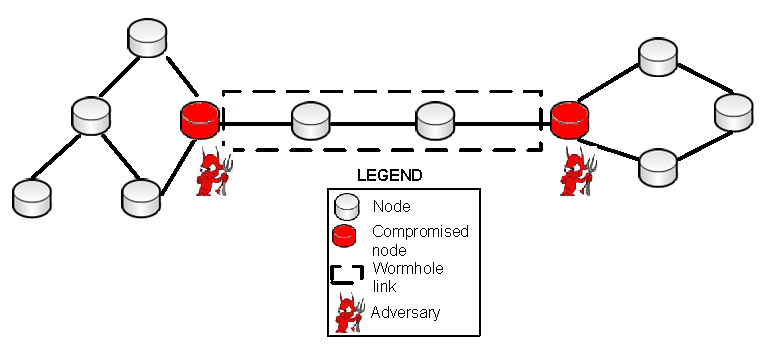} \\
\mbox{(a)} & \mbox{(b)}
\end{array}$
\caption{Illustration of the two classes of wormhole.  (a) In an out-of-band wormhole, the adversary creates a low-latency link between two network regions using a high-capacity channel, such as a directional antenna or wired link.  (b) In an in-band wormhole, the adversary compromises network nodes in different regions and advertises a false one-hop link between  two compromised nodes.  The link actually consists of a path between unsuspecting valid nodes.}
\label{fig:wormhole_illustration}
\end{figure*}
\subsection{Wormhole Attack and Mitigation}
In a wormhole attack, an adversary creates a covert path (referred to as \emph{wormhole tunnel}) that connects two distant regions of the network. Since the wormhole creates the appearance of a short path between distant regions of the network, shortest-path routing protocols will route a large fraction of the network traffic through the wormhole tunnel. The adversary can then control this traffic and selectively drop packets, increase delays, or create routing instability. The wormhole link can also be used to record messages overheard in one network region, such as sensed data or control signals, and replay those messages in order to disrupt the performance of one or more system components.  The wormhole can be further classified as out-of-band or in-band, depending on the nature of the wormhole tunnel.
\subsubsection{Out-of-band wormhole formation}
In the out-of-band wormhole, an adversary establishes a low-latency link (wormhole link) between two distant regions of the network (Figure \ref{fig:wormhole_illustration}(a)). This may be done through wired links that are not available to network nodes, or through high-gain directional wireless antennas. Once the adversary has gained control over a large amount of packets flowing through the wormhole link, the adversary can disrupt the system performance by dropping or delaying packets.
In order to create an out-of-band wormhole, the adversary does not need to compromise any node or cryptographic secrets. 
\subsubsection{Out-of-band wormhole mitigation}
The out-of-band wormhole is based on replaying messages that are intended for a local geographic area in a different geographic region.  As a result of physical constraints on propagation through the medium, the time for a message to propagate to a node's immediate neighbors will be less than the time required for the message to propagate to the eavesdropper, traverse the wormhole tunnel, and then propagate to any nodes on the other side of the wormhole tunnel.  This discrepancy is the basis for the packet leash defense~\cite{hu2003packet}, in which the sender of each packet attaches an expiration time to the packet, equal to $t_{s} + \frac{R}{c} + \Delta$, where $t_{s}$ is the transmission time, $\frac{R}{c}$ is the propagation time, and $\Delta$ is an estimate of the clock skew between the sending and receiving nodes.  All packets received after their expiration time are discarded. Packets are signed using message authentication codes to prevent the adversary from modifying the expiration time.


\subsubsection{In-band wormhole formation}
In the in-band wormhole attack, an adversary compromises two nodes in different regions of the network and falsely advertises a one-hop link between those nodes via the routing protocol. As in the out-of-band case, the appearance of this short path will result in a large traffic flow into the two compromised nodes. The adversary then chooses a path, consisting of both valid and compromised nodes, between the two nodes comprising the wormhole tunnel.  The in-band wormhole requires the adversary to compromise  at least two nodes, but does not require any specialized hardware.  The in-band wormhole is illustrated in Figure \ref{fig:wormhole_illustration}(b).
\subsubsection{In-band wormhole collapse}
\begin{figure}
\centering
$\begin{array}{cc}
\includegraphics[width=3.5in]{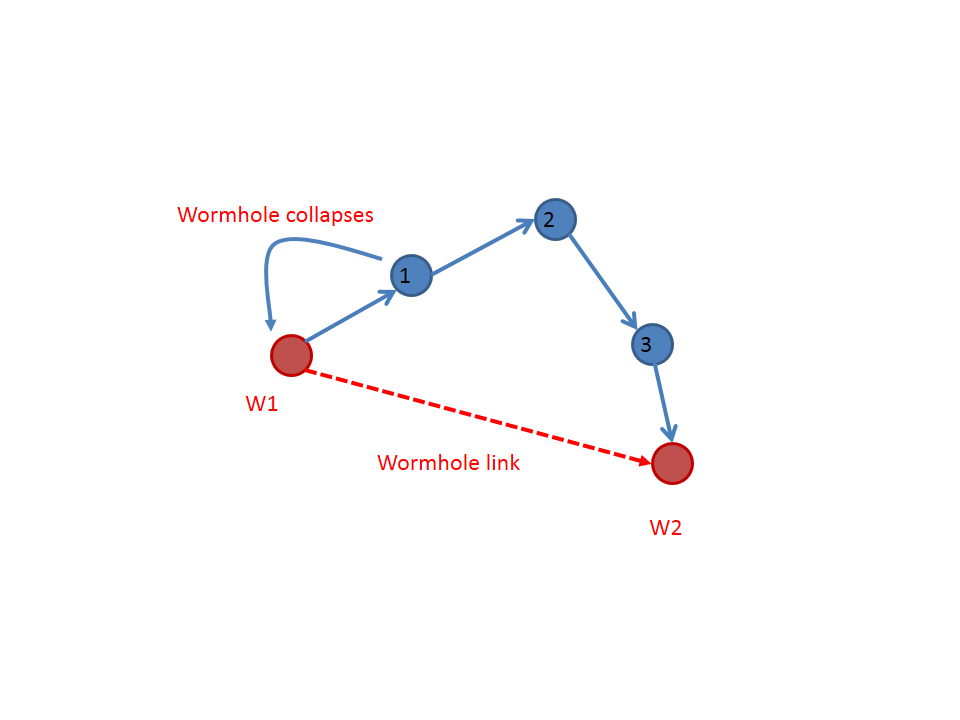} \vspace{-40pt} &
\includegraphics[width=3.5in]{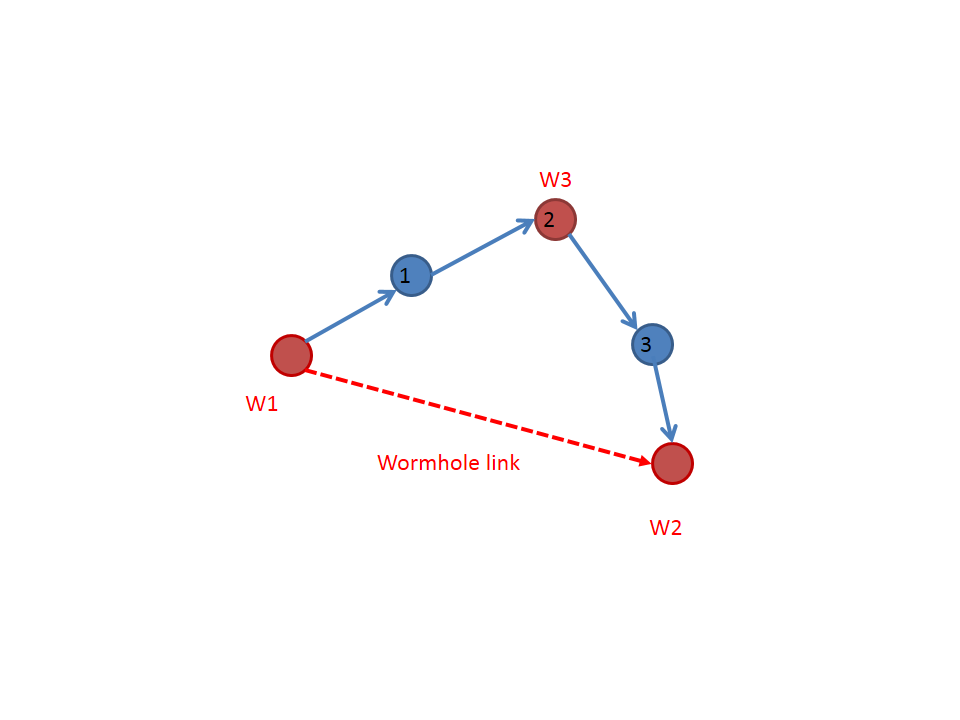} \\
\mbox{(a)} & \mbox{(b)}
\end{array}$
\caption{Illustration of the collapse of in-band wormholes. (a) When the colluding nodes $W_{1}$ and $W_{2}$ advertise a one-hop link between them, the intermediate nodes on the path between $W_{1}$ and $W_{2}$ will attempt to forward packets through the advertised $(W_{1}, W_{2})$ link, creating a routing loop that causes the wormhole to collapse. (b) By tunneling packets to an intermediate node $W_{3}$ satisfying the conditions of Lemma \ref{lemma:collapse}, which then forwards the packets to $W_{2}$, the adversary avoids wormhole collapse.}
\label{fig:wormhole_collapse}
\end{figure}

In order to create an in-band wormhole, the adversary must avoid wormhole collapse, which occurs under the following conditions.  The wormhole tunnel consists of a path between two colluding nodes, denoted $W_{1}$ and $W_{2}$.  The intermediate nodes in the tunnel, however, will attempt to route packets from $W_{1}$ to $W_{2}$ using shortest-path routing.  Since the wormhole tunnel is advertised as a one-hop link between $W_{1}$ and $W_{2}$, any packets sent from $W_{1}$ to $W_{2}$ are likely to be forwarded back to $W_{1}$, creating a routing loop (Figure \ref{fig:wormhole_collapse}(a)).

To avoid wormhole collapse, the adversary must capture a third node, denoted $W_{3}$.  Instead of routing packets directly from $W_{1}$ to $W_{2}$ in the wormhole link, the adversary sends packets from $W_{1}$ to $W_{3}$, and then from $W_{3}$ to $W_{2}$, as shown in Figure \ref{fig:wormhole_collapse}(b).  The conditions on $W_{3}$ to prevent wormhole collapse are given  by the following lemma.
\begin{lemma}[\cite{mahajan2008analysis}]
\label{lemma:collapse}
Let $d(i,j)$ denote the length of the shortest path between nodes $i$ and $j$.  Then the wormhole tunnel formed by colluding nodes $W_{1}$, $W_{2}$, and $W_{3}$ does not collapse if 
\begin{displaymath}
d(W_{1}, W_{3}) < d(W_{2}, W_{3}) + 3.
\end{displaymath}
\end{lemma}
\subsubsection{In-band wormhole mitigation}
Since the in-band wormhole is mounted using compromised nodes and their stored cryptographic keys, defenses against the out-of-band wormhole may be ineffective against in-bandwidth wormholes.  The in-band wormhole, however, will incur longer delays than the out-of-band wormhole, since it relies on a multi-hop path of network nodes to forward packets.  By performing statistical analysis, the network nodes  can identify one-hop links with exceptionally long delays and/or packet-loss rates, which are then suspected of being wormhole links and ignored for routing purposes~\cite{kruus2006band}.  

\section{Proposed Passivity  Framework for Out-of-Band Wormhole}
\label{sec:approach}
In this section, we introduce our passivity-based framework for modeling and mitigating out-of-band wormholes in a networked control system.  Our model considers the effect of the wormhole attack and mitigation on the delay and flow allocation of the network traffic. We first develop a dynamical model for the flow allocation by the network nodes. We then model the delays experienced due to the out-of-band wormhole, followed by the effect of mitigation mechanisms. Lastly, we consider the interconnection of these three dynamical models and  characterize the flow allocation and delay at the unique equilibrium point via a passivity-based approach.

\subsection{Dynamical Model of Network Flow Allocation}  We assume that each source node $S_{i}$ maintains a flow with total rate $r_{i}$ to destination $D_{i}$.  This flow is divided among the paths $\mathcal{P}_{i}$ used by source $S_{i}$ in order to minimize the overall delay.  Let $r_{P}(t)$ denote the flow allocated to path $P \in \mathcal{P}_{i}$ at time $t$, so that $\sum_{P \in \mathcal{P}_{i}}{r_{P}} = r_{i}$. The vector of flow rates is denoted $\mathbf{r}_{i}(t) \triangleq \{r_{P}(t) : P \in \mathcal{P}_{i}\}$. Furthermore, let $f_{l}(r_{l})$ denote the delay experienced on link $l$ when the rate of flow on link $l$ is given by $r_{l}$.  Let $q_{P}(r_{P}) \triangleq \sum_{l \in P}{f_{l}(r_{l})}$ denote the total delay on path $P$, equal to the sum of the delays on each link comprising the path, where $\{l \in P\}$ denotes summing over the links $l$ in path $P$.  Finally, define the $L \times \left(\sum_{i=1}^{n}{m_{i}}\right)$ matrix $A$ by
\begin{displaymath}
A_{lP} = \left\{
\begin{array}{cl}
1, & \mbox{link $l$ in path $P$} \\
0, & \mbox{else}
\end{array}
\right.
\end{displaymath}
so that $r_{l} = (A\mathbf{r})_{l}$.

Achieving the minimum possible delay is equivalent to finding $\{r_{P}: P \in \mathcal{P}_{i}\}$ satisfying
\begin{displaymath}
\min{\left\{\sum_{P \in \mathcal{P}_{i}}{r_{P}q_{P}(r_{P})} : \sum_{P \in \mathcal{P}_{i}}{r_{P}} = r_{i}\right\}},
\end{displaymath}
since $r_{P}q_{P}(r_{P})$ is the total delay on path $P$, $\sum_{P \in \mathcal{P}_{i}}{r_{P}}$ is the overall delay experienced on all paths, and $\sum_{P \in \mathcal{P}_{i}}{r_{P}} = r_{i}$ is a constraint on the total throughput.
Determining whether this condition is satisfied requires the source $S_{i}$ to determine the incremental change in delay from shifting flow from path $P$ to path $P^{\prime}$ for all $P, P^{\prime} \in \mathcal{P}_{i}$.  The incremental change, however, depends on parameters that the source cannot observe, such as the rates of the other sources and the excess capacity of each link, and hence cannot be computed directly by the source. Instead, we assume that each source attempts to minimize the total delay based on the currently observed delay characteristics of each link. This condition is formalized by the concept of a \emph{Wardrop equilibrium} \cite{altman2004equilibrium}, defined as follows.
\begin{definition}
\label{def:Wardrop}
The flow allocation $\{r_{P}: P \in \mathcal{P}_{i}\}$ is a \emph{Wardrop equilibrium} for source $S_{i}$ if for any path $P$, $r_{P} > 0$ implies that $q_{P} \leq q_{P^{\prime}}$ for all $P^{\prime} \in \mathcal{P}_{i}$.
\end{definition}

Definition \ref{def:Wardrop} implies that a positive flow rate is allocated to path $P \in \mathcal{P}_{i}$ if and only if there is no path $P^{\prime}$ currently experiencing lower delays than path $P$.  We now introduce flow rate dynamics that, when used by each source $S_{i}$ to choose $\mathbf{r}_{i}(t)$, cause the network to converge to a Wardrop equilibrium.
We prove convergence to the Wardrop equilibrium by first proving that $\mathbf{r}_{i}$ is a steady state for the dynamics if and only if it is a Wardrop equilibrium, and that the Wardrop equilibrium is unique. We then use a passivity-based approach to prove the system converges to a unique steady state and hence converges to the Wardrop equilibrium.

Let $P_{i}^{min}(q)$ denote a time-varying index satisfying
\begin{displaymath}
P_{i}^{min}(q) \in \arg\min{\left\{q_{P} : P \in \mathcal{P}_{i}\right\}}.
\end{displaymath}
 We define the dynamics of the flow rate $r_{P}(t)$ allocated to path $P \in \mathcal{P}_{i}$ by
\begin{equation}
\label{eq:Wardrop_dynamics}
\dot{r}_{P}(t) = \left\{
\begin{array}{lc}
-\{q_{P}(r_{P}(t)) - q_{P^{min}}(r_{P_{i}^{min}}(t))\}_{+}^{r_{P}}, & P \neq P_{i}^{min}(q) \\
-\sum_{P \neq P_{i}^{min}(q)}{\dot{r}_{P}(t)}, & P = P_{i}^{min}(q)
\end{array}
\right.
\end{equation}
where
\begin{displaymath}
\{x\}_{+}^{r_{P}} = \left\{
\begin{array}{cl}
0, & x > 0 \mbox{ and } r_{P} = 0 \\
x, & \mbox{else}
\end{array}
\right.
\end{displaymath}
Equation (\ref{eq:Wardrop_dynamics}) has the following interpretation.  When the observed delay on path $P$ is greater than the delay observed on path $P^{min}$, which has the minimum delay of any path in $\mathcal{P}_{i}$, the flow allocated to path $P$ is reduced if it is positive.  When the path $P$ has the minimum delay of any path in $\mathcal{P}_{i}$ ($P = P_{i}^{min}(q)$), additional flow is allocated to path $P$ (note that, since $\dot{r}_{P}(t) \leq 0$ if $P \neq P_{i}^{min}(q)$, $-\sum_{P \neq P_{i}^{min}(q)}{\dot{r}_{P}(t)} \geq 0$).  Since the total flow from source $S_{i}$ is constant, the dynamics are chosen such that $\sum_{P \in \mathcal{P}_{i}}{\dot{r}_{P}(t)} = 0$.
The following proposition verifies that the dynamics (\ref{eq:Wardrop_dynamics}) define a feasible flow allocation for all time $t$.
\begin{proposition}
\label{prop:well_defined}
Suppose that $\sum_{P \in \mathcal{P}_{i}}{r_{P}(0)} = r_{i}$ and $r_{P}(0) \geq 0$ for all $P \in \mathcal{P}_{i}$.  Then for all $t > 0$, $\sum_{P \in \mathcal{P}_{i}}{r_{P}(t)} = r_{i}$ and $r_{P}(t) \geq 0$ for all $P \in \mathcal{P}_{i}$.
\end{proposition}

A proof is given in the appendix.  We next show that the equilibria of (\ref{eq:Wardrop_dynamics}) are equivalent to the Wardrop equilibria of the system.
\begin{proposition}
\label{prop:Wardrop_equilibrium}
The dynamics (\ref{eq:Wardrop_dynamics}) have an equilibrium at $\mathbf{r}_{i}^{\ast}$ if and only if $\mathbf{r}_{i}^{\ast}$ is a Wardrop equilibrium.
\end{proposition}
\begin{IEEEproof}
First, suppose that $\dot{r}_{P}(t) = 0$ for all $P \in \mathcal{P}_{i}$, and assume that $\mathbf{r}_{i}(t)$ is not a Wardrop equilibrium.  By Definition \ref{def:Wardrop}, there exists $P$ such that $q_{P}(r_{P}) > q_{P^{min}_{i}(t)}(r_{P^{min}_{i}})$ and $r_{P}(t) > 0$.  The condition $\dot{r}_{P}(t) = 0$ implies that
\begin{equation}
\label{eq:Wardrop_eqbm_proof}
\{q_{P}(r_{P}) - q_{P^{min}_{i}}(r_{P}^{\ast})\}_{+}^{r_{P}} = 0.
\end{equation}
Since $q_{P}(r_{P}) > q_{P^{min}_{i}}(r_{P^{min}_{i}})$, condition (\ref{eq:Wardrop_eqbm_proof}) holds if and only if $r_{P} = 0$, contradicting the assumption that $r_{P} > 0$.

Now, suppose that $\mathbf{r}_{i}$ is a Wardrop equilibrium.  The goal is to show that $\mathbf{r}_{i}$ is an equilibrium of (\ref{eq:Wardrop_dynamics}).  Consider $P \in \mathcal{P}_{i}$, and suppose that $P \neq P^{min}_{i}$.  We show that $\dot{r}_{P}(t) = 0$ by separately considering the cases where $q_{P}(r_{P}) = q_{P^{min}_{i}}(r_{P^{min}_{i}})$ and $q_{P}(r_{P}) > q_{P^{min}_{i}}(r_{P^{min}_{i}})$ ($q_{P}(r_{P}) < q_{P^{min}_{i}}(r_{P^{min}_{i}})$ contradicts the definition of $P^{min}_{i}$).

If $q_{P}(r_{P}) = q_{P^{min}_{i}}(r_{P}^{\ast})$, then $\dot{r}_{P}(t) = 0$.  On the other hand, if $q_{P}(r_{P}) > q_{P^{min}_{i}}(r_{P^{min}_{i}})$, then the delay experienced on path $P$ exceeds the minimum delay, and therefore $r_{P}(t) = 0$ by Definition \ref{def:Wardrop}.  Hence
\begin{displaymath}
\dot{r}_{P}(t) = \{q_{P}(r_{P})-q_{P^{min}_{i}}(r_{P^{min}_{i}})\}_{+}^{r_{P}} = 0.
\end{displaymath}
Finally, we have
\begin{displaymath}
\dot{r}_{P^{min}_{i}} = -\sum_{P \neq P^{min}_{i}}{\dot{r}_{P}} = 0,
\end{displaymath}
which proves that $\mathbf{r}_{i}$ is an equilibrium point of (\ref{eq:Wardrop_dynamics}).
\end{IEEEproof}

Proposition \ref{prop:Wardrop_equilibrium} implies that the equilibria of (\ref{eq:Wardrop_dynamics}) are equal to the Wardrop equilibria of the system. The following Lemma proves the equilibria of (\ref{eq:Wardrop_dynamics}) are unique.
\begin{lemma}
\label{lemma:unique_Wardrop}
If the functions $f_{l}: \mathbb{R} \rightarrow \mathbb{R}$ are strictly increasing for all links $l$, then there exists a unique equilibrium for the dynamics (\ref{eq:Wardrop_dynamics}).
\end{lemma}

\begin{figure}[h]
\centering
\includegraphics[width=4.5in]{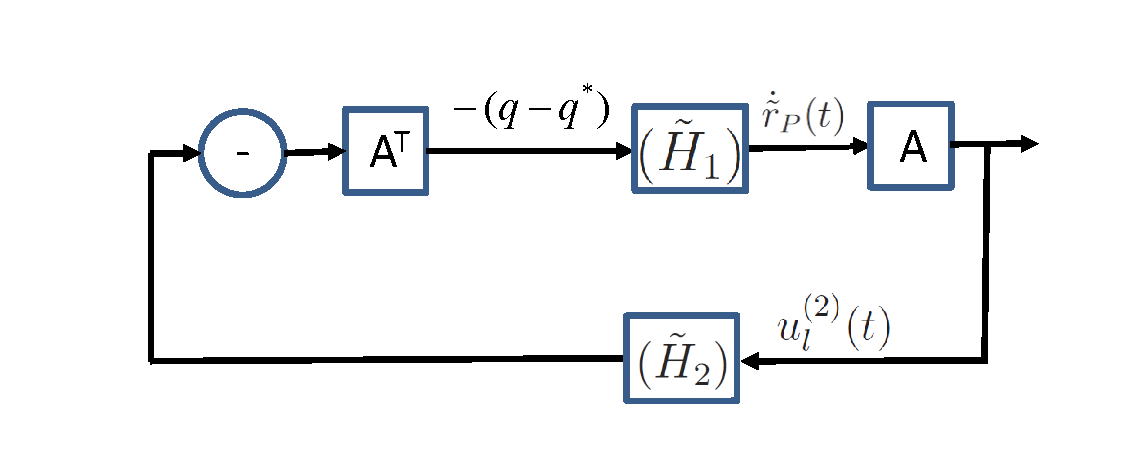}
\vspace{-20pt}
\caption{Illustration of the flow allocation and link delay dynamics $(\tilde{H}_{1})$ and $(\tilde{H}_{2})$. The passive system $(\tilde{H}_{1})$ represents the flow allocation by each source based on the observed delays at each path. The passive system $(\tilde{H}_{2})$ represents the delays experienced at each link as a function of the flows allocated to the link. Since $(\tilde{H}_{1})$ is strictly passive and $(\tilde{H}_{2})$ is passive, the overall system is asymptotically stable (Theorem \ref{theorem:Wardrop_GAS}).}
\label{fig:Wardrop}
\end{figure}

A proof of Lemma \ref{lemma:unique_Wardrop} is given in the appendix.  Finally, we show that the dynamics (\ref{eq:Wardrop_dynamics}) converge to the  unique Wardrop equilibrium.  As a first step, we present an equivalent representation of (\ref{eq:Wardrop_dynamics}).   We define the system $\tilde{H}_{1}$, which takes input $u^{1} \in \mathbb{R}^{m_{i}}$, by
\begin{displaymath}
(\tilde{H}_{1}) \ \left\{
    \begin{array}{ll}
    \dot{\tilde{r}}_{P}(t) = -\{q^{\ast}_{P} - u_{P} - q^{\ast}_{P^{min}(q^{\ast}-u)} + u_{P^{min}(q^{\ast}-u)}\}_{+}^{r_{P}}, & P \neq P_{i}^{min}(q^{\ast}-u) \\
    \dot{\tilde{r}}_{P}(t) = -\sum_{P \neq P^{min}(q^{\ast}-u)}{\dot{\tilde{r}}_{P}(t)}, & P = P_{i}^{min}(q^{\ast}-u) \\
    \tilde{y}_{P}(t) = \dot{\tilde{r}}_{P}(t), & \forall P \in \mathcal{P}_{i}
    \end{array}
    \right.
\end{displaymath}
We define a system $(\tilde{H}_{2})$, which takes input $u^{(2)}(t) \in \mathbb{R}^{L}$, as
\begin{displaymath}
(\tilde{H}_{2}) \ \left\{
\begin{array}{l}
\dot{z}_{l}(t) = u^{(2)}_{l}(t) \\
y_{l}(t) = f_{l}(z_{l}(t)) - f_{l}(z_{l}^{\ast})
\end{array}
\right.
\end{displaymath}
where $z_{l}^{\ast}$ is the rate through link $l$ in the unique equilibrium guaranteed by Lemma \ref{lemma:unique_Wardrop}.  We let $(\tilde{H})$ denote the system formed by a negative feedback interconnection between $(\tilde{H}_{1})$ and $(\tilde{H}_{2})$ (Figure \ref{fig:Wardrop}).  The following proposition establishes the equivalence between the state dynamics defined by (\ref{eq:Wardrop_dynamics}) and system $(\tilde{H})$.  

\begin{proposition}
    \label{prop:neg_equivalent}
    For all $t$, $\tilde{\mathbf{r}}(t) = \mathbf{r}(t)$. 
    \end{proposition}
A proof is given in the appendix.  The following theorem establishes that the flow rate allocation converges to the unique Wardrop equilibrium.


\begin{theorem}
\label{theorem:Wardrop_GAS}
Suppose that $\sum_{P \in \mathcal{P}_{i}}{r_{P}(0)} = r_{i}$.  If the link delay $f_{l}(r_{l})$ is strictly increasing as a function of $r_{l}$ for all links $l$, then
\begin{displaymath}
\lim_{t \rightarrow \infty}{\mathbf{r}_{i}(t)} = \mathbf{r}_{i}^{\ast},
\end{displaymath}
where $\mathbf{r}_{i}^{\ast}$ is the unique Wardrop equilibrium.
\end{theorem}

\begin{IEEEproof}
It suffices to show that the system $(\tilde{H}_{1})$ is strictly passive from input $\tilde{u}_P^{(1)}(t)$ to output $\tilde{y}_{P}^{(1)}(t)$ and the system $(\tilde{H}_{2})$ is passive from input ${u}_{l}^{(2)}(t)$ to output ${y}_{l}^{(2)}(t)$.

    Define the function $V_{1}(\mathbf{r}) = q^{\ast T}(\mathbf{r} - \mathbf{r}^{\ast})$, and let $q = q^{\ast} - u$.  Then $$\dot{V}_{1}(\mathbf{r}) = q^{\ast T}\dot{\mathbf{r}} = q^{T}\dot{\mathbf{r}} + \tilde{u}^{(1)}(t)^{T}\dot{\mathbf{r}}.$$ To prove strict passivity, it therefore suffices to show that $q^{T}\dot{\mathbf{r}} < 0$.  Without loss of generality, suppose that $|\mathcal{P}_{i}| = m_{i} + 1$ and $P_{i}^{\min}(q^{\ast} - u) = P_{i}^{\min}(q) = m_{i}+1$.  Then
    \begin{IEEEeqnarray*}{rCl}
    q^{T}\dot{\mathbf{r}} &=& \sum_{j=1}^{m_{i}+1}{q_{j}\dot{r}_{j}(t)} = \sum_{j=1}^{m_{i}}{q_{j}\dot{r}_{j}(t)} - \sum_{j=1}^{m_i}{q_{m_{i}+1}\dot{r}_{j}(t)} \\
    &=& -\sum_{j=1}^{m_{i}}{q_{j}(q_{j} - q_{m_{i}+1})_{+}^{r_{j}}} + \sum_{j=1}^{m}{q_{m_{i}+1}(q_{j}-q_{m_{i}+1})_{+}^{r_{j}}}.
    \end{IEEEeqnarray*}
    By definition, $(q_{j} - q_{m_{i}+1})_{+}^{r_{j}} \geq 0$.  Furthermore, $q_{j} \geq q_{m_{i}+1}$ for all $j$, and so $$q^{T}\dot{\mathbf{r}} \leq - \sum_{j=1}^{m_{i}}{q_{m_{i}+1}(q_{j}-q_{m_{i}+1})_{+}^{r_{j}}} + \sum_{j=1}^{m_{i}}{q_{m_{i}+1}(q_{j}-q_{m_{i}+1})_{+}^{r_{j}}},$$ thus establishing the passivity of $(\tilde{H}_{1})$.  To prove passivity of $(\tilde{H}_{2})$, define the storage function $$V_{2}(z_{l}) = \int_{0}^{z_{l}-z_{l}^{\ast}}{f_{l}(s + z_{l}^{\ast}) - f_{l}(z_{l}^{\ast}) \ ds}.$$ We have $$\dot{V}_{2}(z_{l}(t)) = (f_{l}(z_{l}(t)) - f_{l}(z_{l}^{\ast}))\dot{z}_{l}(t) = {u}_{l}^{(2)}(t)\tilde{y}_{l}^{(2)}(t),$$ implying passivity of $(\tilde{H}_{2})$.
\end{IEEEproof}

Theorem \ref{theorem:Wardrop_GAS} implies that the flow allocation converges to a unique equilibrium when the delays experienced at each link is a strictly increasing function in flows allocated to each link. The next step in modeling the wormhole attack is to characterize the delays experienced by the wormhole links, which is the topic of the following section.

\subsection{Delay Characteristics of the Out-of-Band Wormhole}
For the out-of-band wormhole, we assume that the wormhole tunnel uses a high-throughput  channel, so that the delay for packets traversing the wormhole tunnel $l$ is equal to the propagation delay $\alpha_{l}$.
Let $\Phi_{l}(r_{l})$ denote the fraction of packets dropped by the wormhole, which we assume to be increasing in $r_{l}$.  The delay for packets traversing the wormhole tunnel, equal to the time per packet transmission multiplied by the average number of retransmissions, is therefore given by
\begin{displaymath}
p_{l} = \frac{\alpha_{l}}{1 - \Phi_{l}(r_{l})}.
\end{displaymath}

Since the packet-loss rate $\Phi_{l}$ is increasing in $r_{l}$, $p_{l}$ is increasing as a function of $r_{l}$ as well, thus preserving the passivity property required by the proof of Theorem \ref{theorem:Wardrop_GAS}.  In what follows, we provide a method for modeling the packet-loss rate $\Phi_{l}(r_{l})$ based on the goals of the adversary.

In mounting the wormhole attack, the goal of the adversary is to attract flow to the wormhole tunnel, in order to either selectively drop packets or mount secondary attacks.  The rate at which packets are dropped by the adversary is equal to $\Phi_{l}(r_{l})r_{l}$, while we model the utility of the adversary from mounting secondary attacks as $U_{A}(r_{l})$.  The adversary's overall utility is therefore given by $\Phi_{l}(r_{l})r_{l} + U_{A}(r_{l})$.  By decreasing $\Phi_{l}$, the adversary increases $r_{l}$ and hence $U_{A}(r_{l})$, at the cost of dropping fewer packets.

The optimal dropping rate depends on the flow rate through the wormhole link in steady-state, which in turn depends on the delays experienced by the other links in the network, since higher delays at other links will increase the flow allocated to the wormhole link.
Based on the network topology, the adversary estimates the delay between source $S_{i}$ and destination $D_{i}$ as $\zeta d(S_{i}, D_{i})$, where $\zeta \geq 0$ is the per-hop delay and $d(\cdot,\cdot)$ is the length of the shortest path between two nodes.  Similarly, the delay experienced by the wormhole path will be equal to $\zeta d(S_{i}, W_{1}) + \frac{\alpha_{l}}{1-\Phi_{l}(r_{l})} + \zeta d(W_{2}, D_{i})$, where $W_{1}$ and $W_{2}$ are the entrance and exit to the wormhole tunnel, respectively.  Define $\Delta_{i,l}$ by
\begin{displaymath}
\Delta_{i,l} = \zeta(d(S_{i}, D_{i}) - (d(S_{i}, W_{1}) + d(W_{2}, D_{i}))).
\end{displaymath}
By Proposition \ref{prop:Wardrop_equilibrium}, the flow from source $S_{i}$ to destination $D_{i}$ will traverse the wormhole tunnel if and only if the delay experienced by the wormhole path is less than the delay experienced by the next-shortest path.  Hence, the flow from source $S_{i}$ to destination $D_{i}$ that traverses the wormhole tunnel in steady-state will be equal to
\begin{displaymath}
r_{i,l}^{\ast} \triangleq \left\{
\begin{array}{cc}
r_{i}, & p_{l} < \Delta_{i,l} \\
0, & \mbox{else}
\end{array}
\right.
\end{displaymath}
Without loss of generality, assume that the indices $i$ are rank-ordered such that $\Delta_{1,l} > \Delta_{2,l} > \cdots > \Delta_{n,l}$, and define $i^{\ast} = \max{\{i : p_{l} < \Delta_{i,l}\}}$.  The flow rate $r_{l}^{\ast}$ traversing the wormhole in steady-state is equal to
\begin{equation}
\label{eq:wormhole_flow}
r_{l}^{\ast} = \sum_{i=1}^{i^{\ast}}{r_{i}}.
\end{equation}
The following proposition describes the set of possible optimal packet-dropping rates $\Phi_{l}^{\ast}$ at equilibrium.
\begin{proposition}
\label{prop:Phi_ast}
The possible solutions $\Phi_{l}^{\ast}$ to the optimization problem
\begin{equation}
\label{eq:adversary_opt}
\begin{array}{cc}
\mbox{maximize} & r_{l}^{\ast}(\Phi_{l})\Phi_{l} + U_{A}(r_{l}^{\ast}(\Phi_{l})) \\
\Phi_{l} & \\
\mbox{s.t.} & \Phi_{l} \in [0,1]
\end{array}
\end{equation}
are given by $\{\gamma_{1}, \ldots, \gamma_{n}\}$, where
\begin{displaymath}
\gamma_{i} = 1 - \frac{\alpha_{l}}{\Delta_{i,l}} - \epsilon
\end{displaymath}
for some $\epsilon << 1$.
\end{proposition}
\begin{IEEEproof}
Suppose that the optimal solution $\Phi_{l}^{\ast}$ to (\ref{eq:adversary_opt}) lies within the interval $(\gamma_{i}, \gamma_{i+1})$ for some $i$. Then by definition of $\gamma_{i}$, $p_{l}^{\ast} > \Delta_{l,i+1}$ and $p_{l}^{\ast} < \Delta_{l,i}$, so that $i^{\ast} = i$. Consider $\Phi_{l}^{\ast} + \delta$ for some $\delta > 0$ satisfying $\Phi_{l}^{\ast} + \delta < \gamma_{i+1}$.  Then by (\ref{eq:wormhole_flow}),
\begin{displaymath}
r_{l}^{\ast}(\Phi_{l}^{\ast}) = \sum_{k=1}^{i}{r_{k}} = r_{l}^{\ast}(\Phi_{l}^{\ast} + \delta).
\end{displaymath}
We therefore have that
\begin{IEEEeqnarray*}{rCl}
r_{l}^{\ast}(\Phi_{l}^{\ast})\Phi_{l}^{\ast} + U_{A}(r_{l}^{\ast}(\Phi_{l}^{\ast})) &<& r_{l}^{\ast}(\Phi_{l}^{\ast})(\Phi_{l}^{\ast} + \delta) + U_{A}(r_{l}^{\ast}(\Phi_{l}^{\ast})) \\
 &=& r_{l}^{\ast}(\Phi_{l}^{\ast} + \delta)(\Phi_{l}^{\ast} + \delta) + U_{A}(r_{l}^{\ast}(\Phi_{l}^{\ast} + \delta)),
 \end{IEEEeqnarray*}
 contradicting the assumption that $\Phi_{l}^{\ast}$ is optimal.
\end{IEEEproof}

The adversary can therefore determine the optimal packet-dropping rate at equilibrium, $\Phi_{l}^{\ast}$, by evaluating $r_{l}^{\ast}\Phi_{l}^{\ast} + U_{A}(r_{l}^{\ast})$ at the set of points $\Phi_{l}^{\ast} = \gamma_{1}, \ldots, \gamma_{n}$ and choosing $\Phi_{l}^{\ast}$ that gives the maximum value of $r_{l}^{\ast}(\Phi_{l})\Phi_{l} + U_{A}(r_{l}^{\ast}(\Phi_{l}))$.  

\subsection{Model of Mitigation for Out-of-Band Wormhole}
The mitigation model is as follows.  Each packet is assigned a packet leash chosen by the source, so that the packet is valid for time $\frac{R}{c} + \Delta_{max}$, where $R$ is the propagation distance, $c$ is the speed of light, and $\Delta_{max}$ is the maximum permissible value of the clock skew.  When the packet traverses a wormhole, the packet violates the packet leash requirement and is dropped when
\begin{displaymath}
\frac{R_{1}}{c} + \frac{R_{2}}{c} + \alpha_{l} + \Delta > \frac{R}{c} + \Delta_{max},
\end{displaymath}
where $R_{1}$ and $R_{2}$ are the distances of the sender and receiver from the wormhole start and end points, respectively, and $\alpha_{l}$ is the wormhole tunnel propagation time as in the previous section. The random variable $\Delta$ represents the clock skew between the nodes comprising the link.  Hence the probability of a packet drop is equal to
\begin{equation}
\label{eq:leash}
P_{d} = \left\{
\begin{array}{ll}
Pr\left(\Delta > \Delta_{max}\right), & l \mbox{ valid} \\
Pr\left(\Delta > \frac{1}{c}(R-R_{1}-R_{2}) - \alpha_{l} + \Delta_{max}\right), & l \mbox{ wormhole}
\end{array}
\right.
\end{equation}
We assume that the network maintains a lower threshold $\Delta_{max}$, representing a more stringent mitigation strategy, when the rate of flow through a link increases.   From (\ref{eq:leash}), the packet drop rate is therefore an increasing function of $\Delta_{max}$.

The effect of the packet leash can be modeled by the increase in delay for each packet due to retransmissions.  This additive delay is equal to $\left(\frac{1}{1-P_{d}}-1\right)f_{l}(r_{l})$, which represents the additional delay due to packet leash. The dynamics of the additive delay introduced by the mitigation mechanism are given as
\begin{displaymath}
(H_{3}) \ \left\{
\begin{array}{l}
\dot{r}_{l}(t) = u_{l}^{(3)}(t) \\
y_{l}^{(3)}(t) = \left(\frac{1}{1-P_{d}}-1\right)(f_{l}(r_{l}(t)))
\end{array}
\right.
\end{displaymath}

\subsection{Steady-state and Stability Analysis for the Out-of-Band Wormhole}
\begin{figure}[h]
\centering
\includegraphics[width=4.5in]{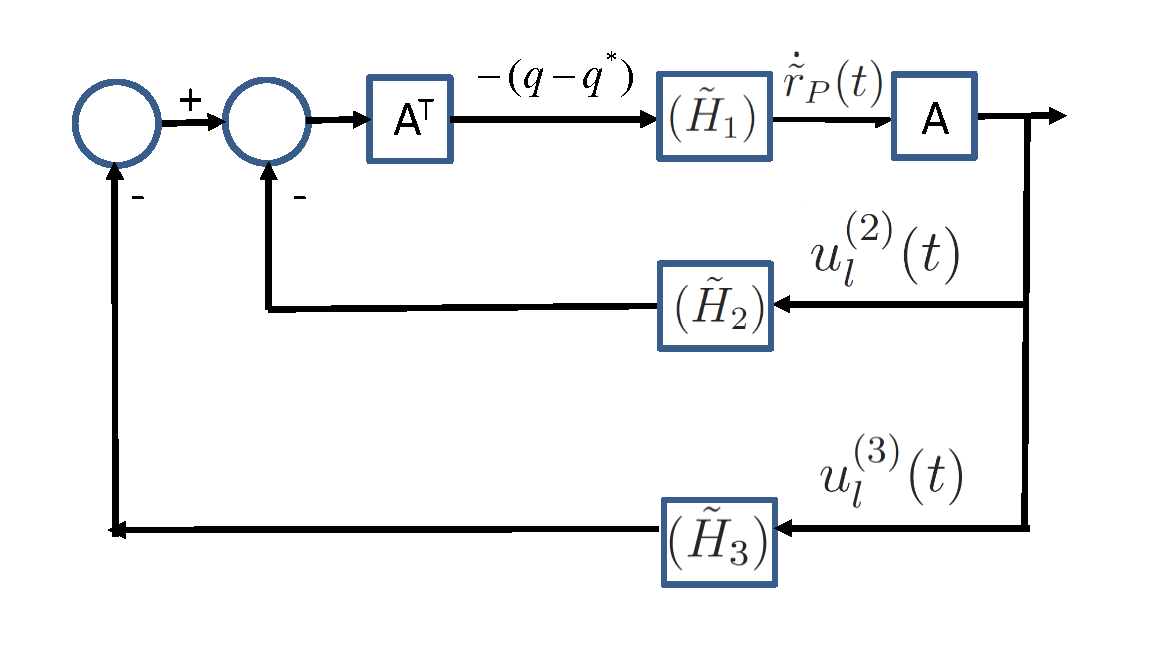}
\vspace{-20pt}
\caption{Block diagram illustrating the out-of-band wormhole link and mitigation. As in Figure \ref{fig:Wardrop}, systems $(\tilde{H}_{1})$ and $(\tilde{H}_{2})$ are passive dynamical systems representing flow allocation and link delays respectively. Passive dynamical system $(\tilde{H}_{3})$ represents the network mitigation mechanisms. By Corollary \ref{corollary:detection_GAS}, the interconnection of these passive systems is asymptotically stable.}
\label{fig:wormhole_detection}
\end{figure}

In this section, we analyze the steady-state characteristics of the overall system.  As a first step, we define the system $(\tilde{H}_{3})$ as
\begin{displaymath}
(\tilde{H}_{3}) \ \left\{
\begin{array}{l}
\dot{r}_{l}(t) = u_{l}^{(3)}(t) \\
\tilde{y}_{l}^{(3)}(t) = \left(\frac{1}{1-P_{d}}-1\right)(f_{l}(r_{l}(t))) - \left(\frac{1}{1 - P_{d}^{\ast}}-1\right)f_{l}(r_{l}^{\ast})
\end{array}
\right.
\end{displaymath}
where $P_{d}^{\ast}$ is the probability of packet drops when the flow allocated to link $l$ is $r_{l}^{\ast}$.
The joint dynamics of the flow allocation, wormhole delay, delays on valid links, and delay introduced by mitigation mechanisms can be represented as a negative feedback interconnection between dynamical systems $(\tilde{H}_{1})$, $(\tilde{H}_{2})$, and $(\tilde{H}_{3})$ (Figure \ref{fig:wormhole_detection}).     The following lemma guarantees global asymptotic stability of the overall system. 

\begin{lemma}
\label{lemma:leash_passive}
The system $(\tilde{H}_{3})$ is passive from input $u_{l}^{(3)}(t)$ to output $\tilde{y}_{l}^{(3)}(t)$.
\end{lemma}
\begin{IEEEproof}
Define $V_{l}(r_{l}) = \int_{r_{l}^{\ast}}^{r_{l}}{\left(\left(\frac{1}{1-P_{d}(s)} - 1 \right) f_{l}(s) - \left(\frac{1}{1-P_{d}^{\ast}}-1\right)f_{l}(r_{l}^{\ast})\right) \ ds}$.

Since
$\left(\frac{1}{1-P_{d}(s)}-1\right)f_{l}(s)$ is nondecreasing as a function of $s$, $V_{l} \geq 0$.  Furthermore, $V_{l}(r_{l}^{\ast}) = 0$ and
\begin{displaymath}
\dot{V}_{l}(t) = \left(\left(\frac{1}{1-P_{d}(r_{l})}-1\right)f_{l}(r_{l}) - \left( \frac{1}{1-P_{d}^{\ast}}-1\right)f_{l}(r_{l}^{\ast})\right)\dot{r}_{l} = \dot{r}_{l}\tilde{y}_{l}^{(3)}(t),
\end{displaymath}
thus establishing passivity of $(\tilde{H}_{3})$.
\end{IEEEproof}

\begin{corollary}
\label{corollary:detection_GAS}
The system of Figure \ref{fig:wormhole_detection} is globally asymptotically stable.
\end{corollary}

\begin{IEEEproof}
By Theorem \ref{theorem:Wardrop_GAS}, the blocks $(\tilde{H}_{1})$ and $(\tilde{H}_{2})$ in Figure \ref{fig:wormhole_detection} form a negative feedback interconnection of passive systems, and hence are passive.  Since $(\tilde{H}_{3})$ is passive by Lemma \ref{lemma:leash_passive}, the system consists of a negative feedback interconnection of passive systems, which is globally asymptotically stable.
\end{IEEEproof}

Corollary \ref{corollary:detection_GAS} implies that the overall system consisting of the flow allocation, out-of-band wormhole, and mitigation converges to a unique stable equilibrium point. This enables the characterization of delay experienced by the networked control system in steady state.

Our passivity based approach for modeling and mitigating in-band wormhole attacks is described in the following section.

\section{Proposed Passivity Framework for In-band Wormhole}
\label{sec:in_band}
In this section, we present a passivity framework for modeling and detecting in-band wormhole attacks mounted by colluding malicious nodes.  As in the out-of-band case, the goal of each source is to select the flow rate on each path in order to minimize the average delay experienced while avoiding the wormhole tunnel. In designing network dynamics, including the source rates and detection mechanism, that achieve this goal, we first model the delay experienced on the wormhole link as a function of the number of compromised nodes. Since the delay depends on the number of compromised nodes, we then model the temporal dynamics of the number of compromised nodes.
 Lastly, we incorporate the impact of the  detection mechanism described in Section \ref{sec:model} on both the valid and wormhole links, and show stability of the overall system.

 \subsection{Delay Characteristics of the In-Band Wormhole}
 \label{subsec:IB_wormhole_price}
 Delays experienced by packets traversing an in-band wormhole are proportional to the number of nodes comprising the wormhole tunnel.  Let $W_{1}$ and $W_{2}$ denote the compromised nodes that create the in-band wormhole tunnel.
 Recall from Section \ref{sec:model} that, in order to avoid wormhole tunnel collapse, packets entering the wormhole tunnel must be routed through a third colluding node, denoted $W_{3}$.
 The number of hops in the wormhole tunnel is therefore equal to $d(W_{1}, W_{3}) + d(W_{3}, W_{2})$.  Furthermore, from Lemma \ref{lemma:collapse}, the node $W_{3}$ must satisfy
 \begin{equation}
 \label{eq:wormhole_stability}
 d(W_{1}, W_{3}) < d(W_{2}, W_{3}) + 3.
 \end{equation}
While the locations of $W_{1}$ and $W_{2}$ are fixed for a given wormhole tunnel, the location of $W_{3}$ depends on the set of nodes compromised by the adversary, denoted $\mathcal{C}$.

In order to minimize delays, and therefore attract more network flow to the wormhole tunnel, the adversary selects the node $W_{3} \in \mathcal{C}$ that minimizes $d(W_{1}, W_{3}) + d(W_{3}, W_{2})$ to collude in establishing the wormhole, subject to the constraint (\ref{eq:wormhole_stability}).
 Letting $x$ denote the fraction of nodes that are misbehaving, and letting $\tilde{\mathcal{C}} = \{W_{3} \in \mathcal{C}: d(W_{1}, W_{3}) < d(W_{2}, W_{3}) + 3\}$, we define
\begin{displaymath}
\beta(x) \triangleq \mathbf{E}[\min{\{d(W_{1},W_{3}) + d(W_{3},W_{2}) : W_{3} \in \tilde{\mathcal{C}}\}} \ | \ |\mathcal{C}| = nx],
\end{displaymath}
where $\mathbf{E}(\cdot)$ denotes expectation and $n$ is the total number of nodes.

Since the delay experienced by the in-band wormhole is a function of the fraction of compromised nodes, a dynamical model of the fraction of compromised nodes is required.

 \subsection{Dynamics of Fraction of Compromised Nodes}
 \label{subsec:IB_comp_dynamics}
The goal of the adversary is to minimize the delay of the wormhole link by compromising nodes.  We let $c_{A}x$, where $c_{A} > 0$, denote the cost of compromising a fraction $x$ of the nodes, and define the adversary's utility function by
\begin{displaymath}
U_{A}(x) \triangleq (n-\beta(x)) -c_{A}x,
\end{displaymath}
where the first term is the reduction in the path length caused by compromising the fraction of nodes $x$, and $c_{A}x$ is the cost.  
In order to obtain the maximum value, we assume that the adversary chooses the rate at which nodes are compromised via a gradient ascent algorithm, so that
\begin{equation}
\dot{x}(t) = (-\beta^{\prime}(x) - c_{A})_{+},
\label{descent}
\end{equation}
where $(z)_{+} = z$ if $z \geq 0$ and $(z)_{+} = 0$ otherwise.
The following proposition proves that $U_{A}(x)$ has a unique global maximum.
\begin{proposition}
\label{prop:beta_concave}
Suppose that, for a given value of $x$, the set of compromised nodes is chosen uniformly at random from $\mathcal{C}_{x} = \{\mathcal{C}: |\mathcal{C}| = nx\}$.  Then the function $\beta(x)$ is decreasing and convex in $x$.
\end{proposition}
A proof is given in the appendix.  Intuitively, $\beta(x)$ is a non-increasing function in $x$ since as the number of colluding malicious nodes increases, the number of paths that can potentially be used as in-band wormhole tunnels also increases.

Proposition \ref{prop:beta_concave} implies that the dynamics (\ref{descent}) converge to a unique equilibrium which is the global maximum of the utility function. To complete the model of the in-band wormhole, the next step is modeling the mitigation by the network.

 \subsection{Model of Mitigation for In-Band Wormhole}
 \label{subsec:IB_detection_model}
 The detection of the in-band wormhole is based on the probability that a communication link is a wormhole tunnel, given observation of the flow rate through the link and the associated delay characteristics.  A link experiencing anomalously long delays is judged to have a high probability of being a wormhole.  We define $B_{1}$ as the event that link $l$ is a wormhole and $B_{0}$ as the event that link $l$ is valid.  Furthermore, we let $w_{l}(t)$ denote the system's belief at time $t$ that the link $l$ is a wormhole, with
 \begin{displaymath}
 w_{l}(t) = Pr(B_{1}|r_{l}(t), p_{l}(t)),
 \end{displaymath}
 where
 \begin{displaymath}
 p_{l}(t) = \left\{
 \begin{array}{ll}
 f_{l}(r_{l}(t)), & \mbox{$l$ valid} \\
 \beta_{l}(x)f(r_{l}), & \mbox{$l$ wormhole}
 \end{array}
 \right.
 \end{displaymath}
 The effect of the detection process on the flow allocation is modeled as an increase in the link price, so that the price is increased by $K\mathbf{1}(w_{l}(t) > \overline{w})$, where $K$ represents a penalty for routing packets through suspected wormhole links, $\mathbf{1}$ denotes the indicator function, and $\overline{w}$ is a predefined threshold.

 A model of the wormhole delay dynamics, taking the derivative of the source rate $\dot{r}$ as input and giving the delay $p-p^{\ast}$ as output, is given by
 \begin{displaymath}
 (H_{l}) \left\{
 \begin{array}{rcl}
 \dot{r}_{l}(t) &=& u(t) \\
 \dot{x}(t) &=& (-\beta^{\prime}(x)-c_{A})_{+} \\
 y_{l}(t) &=& \beta_{l}(x)f(r_{l}) + K(\mathbf{1}(w_{l}(t) > \overline{w})) 
 \end{array}
 \right.
 \end{displaymath}

 The source rate dynamics are unchanged from Section \ref{sec:approach}, since detection is performed at the link instead of the source level.  The steady-state behavior of the system is described as follows.

 \subsection{Steady-state and Stability Analysis for the In-Band Wormhole}
 \label{subsec:IB_stability}
 In this section, we prove the stability of the in-band wormhole, enabling us to characterize the average delay due to the wormhole in steady state.
 Stability of the network in the presence of the in-band wormhole is a result of the following proposition, which establishes the passivity of the wormhole link price.
 \begin{proposition}
 \label{prop:inband_passive}
 The wormhole link dynamics $(H_{l})$ are passive with input $\dot{r}_{l}$ and output $y_{l}$. 
 \end{proposition}
 \begin{IEEEproof}
 To prove passivity when $l$ is a wormhole link, we use the Lyapunov function $V_{l}(\cdot)$ defined by
 \begin{IEEEeqnarray*}{rCl}
 V_{l}(r_{l},x) &=& \int_{r_{l}^{\ast}}^{r_{l}} {\beta_{l}(x)f(s) - \beta_{l}(x^{\ast})f(r_{l}^{\ast}) + K(\mathbf{1}(w_{l}(t) > \overline{w}) - \mathbf{1}(w_{l}^{\ast} > \overline{w})) \ ds} \\
 && + \int_{x^{\ast}}^{x}{\left(\int_{0}^{r_{l}^{\ast}}{f(v) \ dv}\right)\beta_{l}^{\prime}(s) \ ds}.
 \end{IEEEeqnarray*}
 We have
 \begin{IEEEeqnarray*}{rCl}
 \dot{V}_{l}(r_{l},x) &=&  (\beta_{l}(x)f(r_{l}) - \beta_{l}(x^{\ast})f(r_{l}^{\ast}) + K(\mathbf{1}(w_{l}(t) > \overline{w}) - \mathbf{1}(w_{l}^{\ast} > \overline{w})))u_{l} \\
  && + \beta_{l}^{\prime}(x)\left(\int_{r_{l}^{\ast}}^{r_{l}}{f(s) \ ds}  + \int_{0}^{r_{l}^{\ast}}{f(s) \ ds}\right)\dot{x} \\
 &=& y_{l}u_{l} + \beta_{l}^{\prime}(x)\left(\int_{0}^{r_{l}}{f(s) \ ds}\right)(-\beta_{l}^{\prime}(x) - c_{A})_{+} \\
 &\leq& y_{l}u_{l},
 \end{IEEEeqnarray*}
 where the final inequality follows from the the fact that $\beta_{l}(x)$ is nonincreasing (Proposition \ref{prop:beta_concave}) and $f(s) \geq 0$.  The fact that $V_{l}(r_{l}^{\ast},x^{\ast}) = 0$ holds by inspection.  It remains to show that $V_{l}(r_{l}, x) \geq 0$ for all $r_{l}$ and $x$.  This holds because $f_{l}$ and $\mathbf{1}(w_{l}(t) > \overline{w})$ are assumed to be nondecreasing functions of $r_{l}$, while $\beta_{l}^{\prime}$ is an increasing funtion of $x$ by Proposition \ref{prop:beta_concave}.
 \end{IEEEproof}

The stability of the system under the in-band wormhole attack is established by the following theorem.
\begin{theorem}
\label{theorem:in_band_GAS}
The source rate $\mathbf{r}_{i}$ satisfies $\lim_{t \rightarrow \infty}{\mathbf{r}_{i}(t)} = \mathbf{r}_{i}^{\ast}$.
\end{theorem}
\begin{IEEEproof}
The proof follows from the passivity of the source rate (Theorem \ref{theorem:Wardrop_GAS}) and wormhole delay (Proposition \ref{prop:inband_passive}), and the fact that they form a negative feedback interconnection.
\end{IEEEproof}

Theorem \ref{theorem:in_band_GAS} implies that the average delay converges to a stable point in the presence of in-band wormhole.

In what follows, using our framework, we show how complex wormhole attacks consisting of both in- and out-of band wormholes can be jointly modeled and mitigated.

\section{Joint Modeling of  Out-of-Band and In-Band Wormholes}
\label{sec:joint}
At present, in the security literature, out-of-band wormholes and in-band-wormholes are treated using different methods. A more general wormhole that consists of in-band and out-of-band wormholes has not been identified or discussed, though such wormholes can be conceived. Our framework can naturally model complex wormholes formed by composing in-band and out-of-band wormholes.

 We consider a system with a set of out-of-band wormhole links $\mathcal{L}^{\prime} = \{l_{1}^{\prime}, \ldots, l_{w}^{\prime}\}$ and in-band wormhole links $\mathcal{L}^{\prime\prime} = \{l_{1}^{\prime\prime}, \ldots, l_{w}^{\prime\prime}\}$. The delay experienced by a valid link is an increasing function of $r_{l}$, the flow through the link. The delay experienced by an out-of-band wormhole link is defined by the propagation time and the packet dropping rate, as described in Section \ref{sec:approach}. For an in-band wormhole link, the delay experienced by the wormhole link is function of the expected number of hops in the wormhole tunnel, as described in Section \ref{sec:in_band}.
These delay characteristics are described by the following dynamics, where the input $u(t)$ is equal to the change in the source rate $\dot{r}(t)$:
\begin{displaymath}
(H_{l}) \left\{
\begin{array}{ll}
\dot{r}_{l}(t) = u_{l}(t) & \\
y_{l}(t) = \frac{\alpha_{l}}{1-\Phi_{l}(r_{l})} - \frac{\alpha_{l}}{1 - \Phi_{l}(r_{l}^{\ast})}, & l \in \mathcal{L}^{\prime} \\
y_{l}(t) = \beta_{l}(x)f(r_{l}) - \beta_{l}(x^{\ast})f(r_{l}^{\ast}), & l \in \mathcal{L}^{\prime\prime} \\
y_{l}(t) = f_{l}(r_{l}) - f_{l}(r_{l}^{\ast}), & \mbox{else}
\end{array}
\right.
\end{displaymath}
We assume that the network employs mitigation schemes for both the in- and out-of-band wormholes. The out-of-band wormhole mitigation mechanism increases the delay on valid and out-of-band wormhole links as discussed in Section \ref{sec:approach}. However, since the in-band wormhole contains colluding nodes that can modify the time stamps using valid cryptographic keys, the out-of-band wormhole mitigation is ineffective and hence adds no delay to in-band wormhole links. The in-band wormhole mitigation mechanism described in Section \ref{sec:in_band} is employed on the valid and in-band wormhole links. Since the out-of-band wormhole link is created using a high capacity, low-latency channel, the adversary can manipulate the delays in order to thwart the statistical mitigation mechanism.
Hence our model of the impact of mitigation on the link delays is given by the following dynamics:
\begin{displaymath}
(H_{D}) \left\{
\begin{array}{ll}
\dot{r}_{l}(t) = u_{l}(t) & \\
y_{l}(t) = \left(\frac{1}{1-P_{d}}-1\right)f_{l}(r_{l}) - \left(\frac{1}{1-P_{d}^{\ast}} -1\right)f_{l}(r_{l}^{\ast}), & l \in \mathcal{L}^{\prime} \\
y_{l}(t) = K(\mathbf{1}(w_{l}(t) > \overline{w}) - \mathbf{1}(w_{l}^{\ast} > \overline{w})),& l \in \mathcal{L}^{\prime\prime} \\
y_{l}(t) = \left(\frac{1}{1-P_{d}}-1\right)f_{l}(r_{l}) - \left(\frac{1}{1-P_{d}^{\ast}}-1\right)f_{l}(r_{l}^{\ast}) + K(\mathbf{1}(w_{l}(t) > \overline{w}) - \mathbf{1}(w_{l}^{\ast} > \overline{w})), & \mbox{else}
\end{array}
\right.
\end{displaymath}
\begin{figure}
\centering
\includegraphics[width=4.5in]{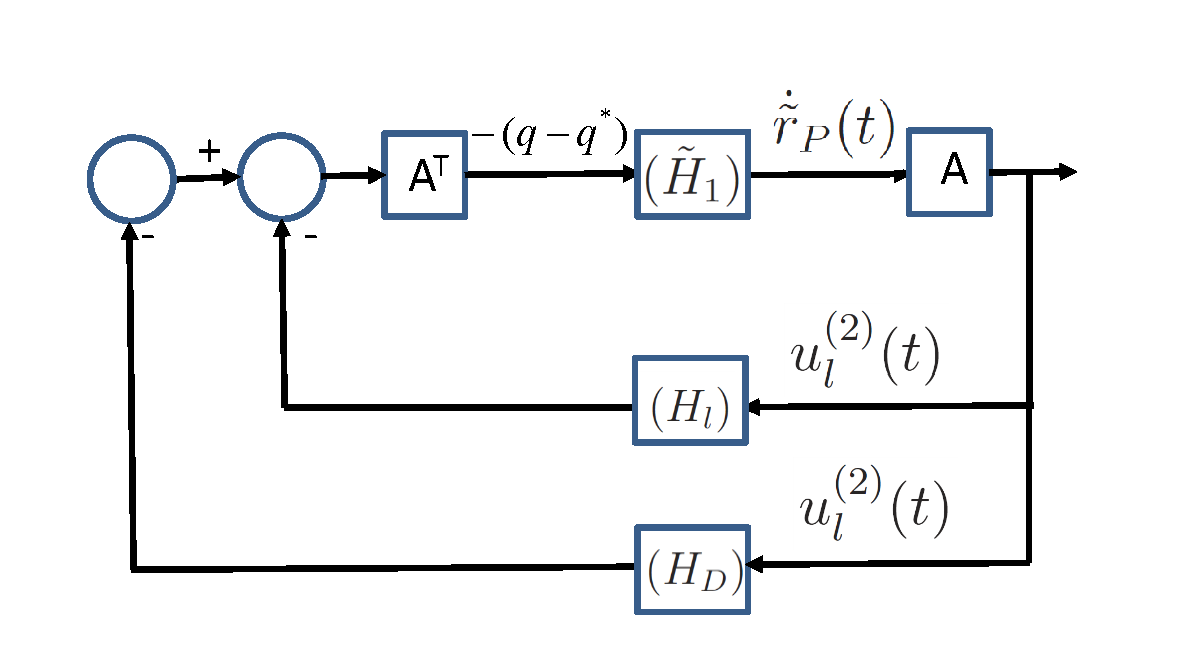}
\vspace{-20pt}
\caption{Illustration of the interconnection between flow allocation, link delay characteristics, and mitigation when multiple out-of-band and in-band wormholes are present. The system $H_{1}$ defines the flow allocation dynamics as a function of observed delays. The system $H_{l}$ defines the delays experienced by valid, out-of-band, and in-band wormhole links as a function of the flow rates. The system $H_{D}$ models the impact of mitigation mechanisms on the flow allocation dynamics.}
\label{fig:joint}
\end{figure}
The flow allocation, delay, and mitigation models are illustrated in Figure \ref{fig:joint}.
The following theorem characterizes the stability properties of the system when both in-band and out-of-band wormholes are present.
\begin{theorem}
\label{theorem:joint_stability}
The interconnected system of Figure \ref{fig:joint} is globally asymptotically stable.
\end{theorem}
\begin{IEEEproof}
By Theorem \ref{theorem:Wardrop_GAS}, the top block of Figure \ref{fig:joint} is strictly passive.  The blocks $H_{l}$ and $H_{D}$  are passive by Theorem \ref{theorem:Wardrop_GAS} (if $l$ is an out-of-band wormhole) and Proposition \ref{prop:inband_passive} (if $l$ is an in-band wormhole).  Hence the negative feedback interconnection of Figure \ref{fig:joint} is globally asymptotically stable.
\end{IEEEproof}

Theorem \ref{theorem:joint_stability} implies the passivity based framework enables us to compose in-band and out-of-band wormholes and characterize the overall delay and flow allocation.

\section{Numerical Study}
\label{sec:simulation}
In this section, we conduct a numerical study using MATLAB. We use our passivity-based framework to answer the following questions for the out-of-band and in-band wormhole attacks: 1) What is the overall flow allocated and delays experienced by sources for a given adversary's strategy? and 2) How do the proposed mitigation methods affect the flow allocation and delays experienced by sources? 3) What is the impact of wormhole attack on a networked control system?

\begin{figure}[h]
\centering
\includegraphics[width=4in]{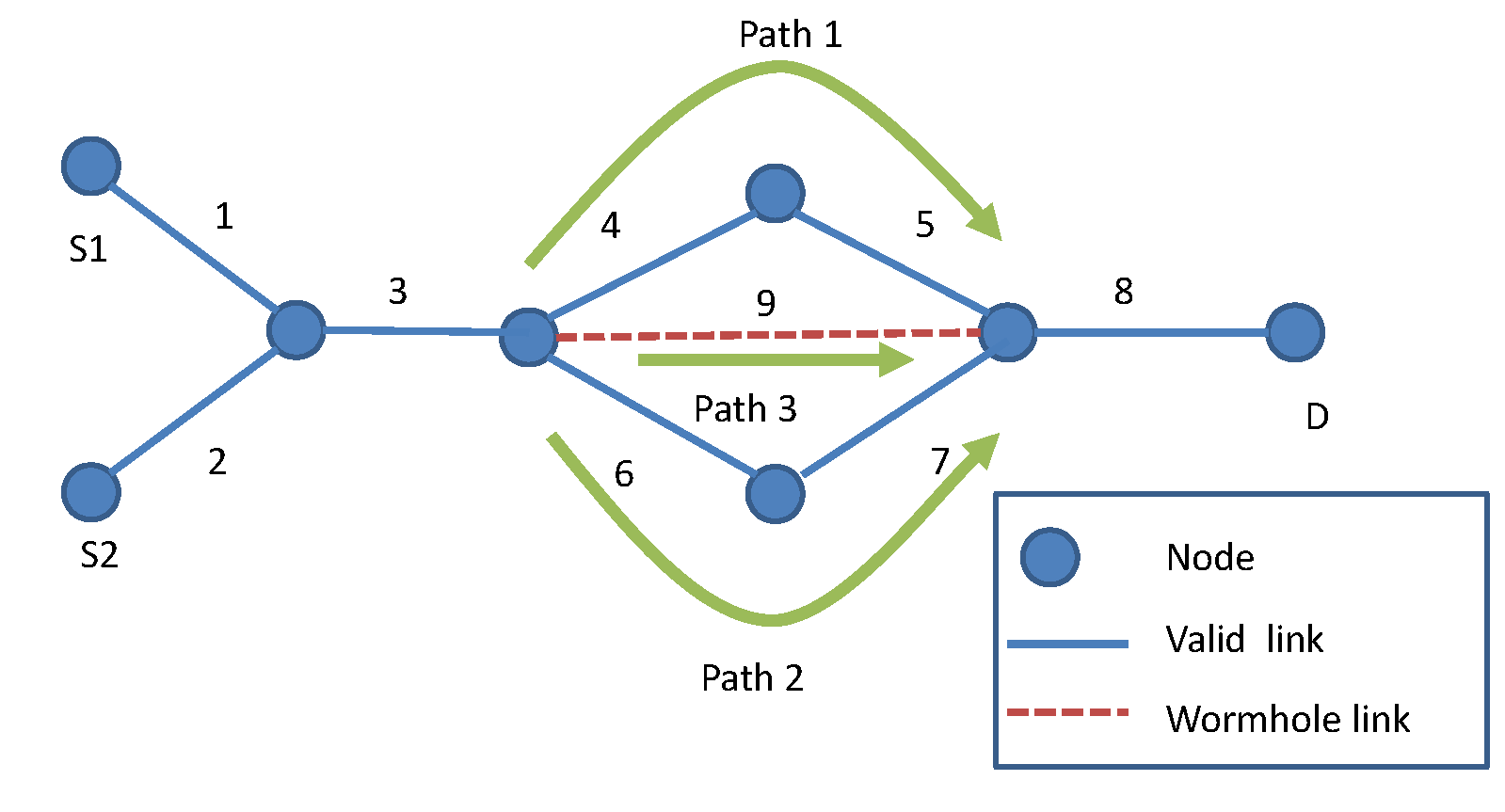}
\caption{Network topology used in numerical study.  Two sources send flows  with total rate of 10 and 5 to destination $D$.  Each source maintains a path through links 4 and 5 (path 1), a path through links 6 and 7 (path 2), and a path through the wormhole link 9 (path 3).}
\label{fig:network}
\end{figure}

We consider a network which consists of two source nodes $S_{1}, S_{2}$ and a single destination node $D$ shown in Figure \ref{fig:network}. The source rates for sources 1 and 2 are given as 10 and 5, respectively. Each source allocates flows to three different paths. We denote the path which traverses  links 4 and 5 as path 1, the path which traverses links 6 and 7 as path 2, and the path which traverses link 9 as path 3.
We assume the propagation delays for valid links are equal and normalized to 1 time unit. The average delay is given as the propagation delay times the expected number of transmissions. The expected number of transmissions for a link is given as $\frac{1}{1 - P_{d}}$ where $P_{d}$ is the probability of a packet drop. For valid links, we assume the probability of packet drop is due to buffer overflow in an M/M/1/K queue \cite{ross2009introduction}. We denote $\rho_{l} = \frac{r_{l}}{c_{l}}$, where $r_{l}$ is the amount of traffic flowing into link $l$, and $c_{l}$ is the capacity of link $l$. The probability of a packet drop is given as
\begin{equation}
P_{d} = \frac{\rho^{K} - \rho^{K+1}}{1-\rho^{K+1}}
\end{equation}
In this simulation, we assume the buffer size $K=5$ for all links.

The propagation delay for the wormhole tunnel, $\alpha_{l}$, is assumed to be 2. The clock skew $\Delta$ is an exponential random variable with mean 1. $\Delta_{\max}$ is given as
\begin{equation}
\Delta_{\max} = \alpha_{l}-1 + \frac{1}{r},
\end{equation}
which is a monotonic decreasing function in $r$.
\begin{figure*}[t]
\centering
$\begin{array}{cc}
\includegraphics[width=2.5in]{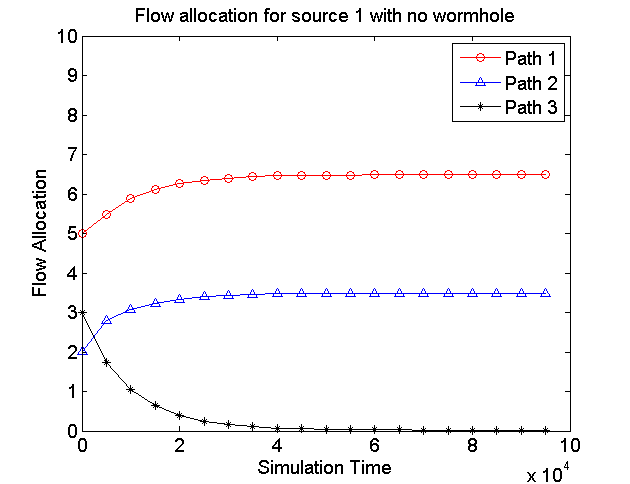} &
\includegraphics[width=2.5in]{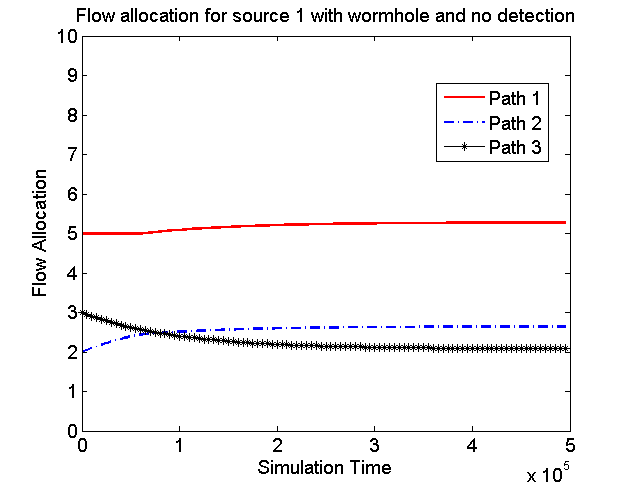} \\
\mbox{(a)} & \mbox{(b)} \\
\includegraphics[width=2.5in]{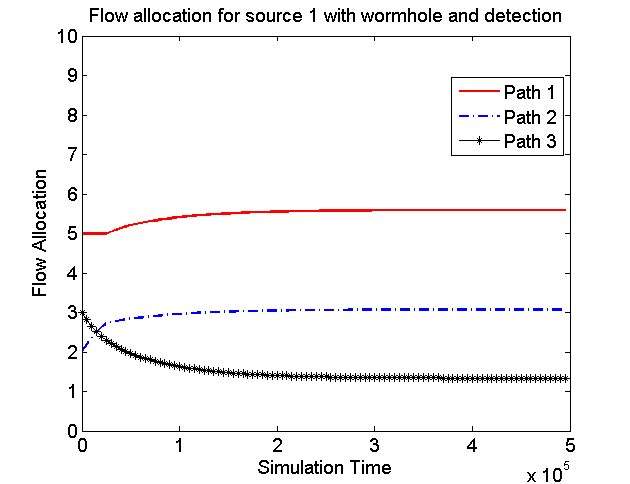} &
\includegraphics[width=2.5in]{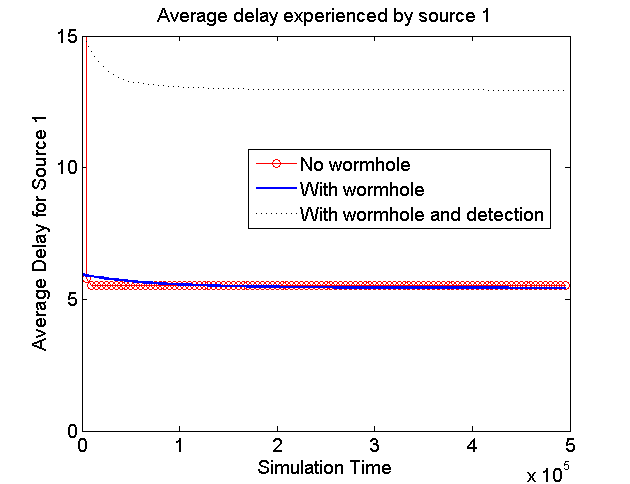} \\
\mbox{(c)} & \mbox{(d)}
\end{array}$
\caption{Simulation of our passivity framework for modeling the out-of-band wormhole. The time scales represent the number of iterations of the simulation. Each iteration represents a single update step of the wormhole dynamics. The source rates for sources 1 and 2 are given as 10 and 5. Initial flow allocation for source 1 is [5,2,3], and flow allocation for source 2 is [2,2,1] for paths 1, 2, and 3 respectively. (a) The convergence of flow allocation without the wormhole when link 9 has capacity 0.01. (b) The impact of the wormhole on flow allocation with no mitigation mechanisms. (c) The flow allocation when packet leashes method are used.}
\label{outofband}
\end{figure*}

\subsection{Simulation of the Out-of-band Wormhole}
In the out-of-band wormhole simulation, we assume that link 9 in Figure \ref{fig:network} is the wormhole link. The propagation delay for the wormhole link is denoted as $\alpha_l$, where $\alpha_l > 1$ due to longer physical distance the packet traverses through the wormhole link. The delay for the wormhole link is given as $\frac{\alpha_l}{1 - \Phi(r)}$ where $\Phi(r)$ is the dropping rate of packets that flow into the wormhole link. The function $\Phi(r)$ is given as $\Phi(r) =(1-\frac{1}{r}) \mathbf{1}_{(r >1)}$.

We illustrate the impact of the out-of-band wormhole  by comparing the flow allocation without the wormhole (Figure \ref{outofband}(a)) with the flow allocation resulting from the wormhole (Figure \ref{outofband}(b)). In both cases, simulation result shows that our choice of dynamics results in the convergence to the stable equilibrium. Figure \ref{outofband}(a) shows that when path 3 contains a poor quality link with low capacity of 0.01 (link 9), both sources allocate negligible amount of flows to path 3 in equilibrium.
Figure \ref{outofband}(b) shows that packet drops by the wormhole path result in increased delay on path 3, thus reducing the flow allocated to path 3. At the equilibrium, the wormhole drops half of the packets on average. As a result, the wormhole is able to attract only 2 units of flow from both source 1 and 2 combined.

Figure \ref{outofband}(d) shows that in order to attract flow, the wormhole has to provide a low-latency link whose performance is comparable to other links. Therefore, the average delay experienced by the sources is approximately the same regardless of the presence of the wormhole link.
Figure \ref{outofband}(c) shows that when packet leash mitigation methods are  employed, the amount of flow allocated to the wormhole link is reduced from 2 to 1.3 units.  The overall delay, however, increases due to packet drops caused by the packet leash mitigation method.

\subsection{Simulation of the In-band Wormhole}
\begin{figure*}[t]
\centering
$\begin{array}{ccc}
\includegraphics[width=2.1in]{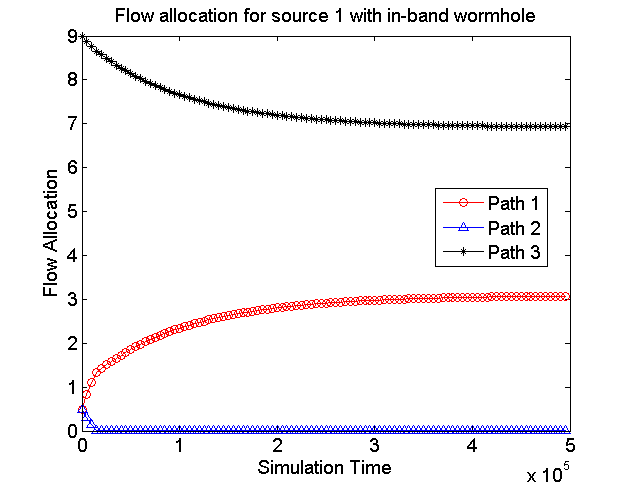} &
\includegraphics[width=2.1in]{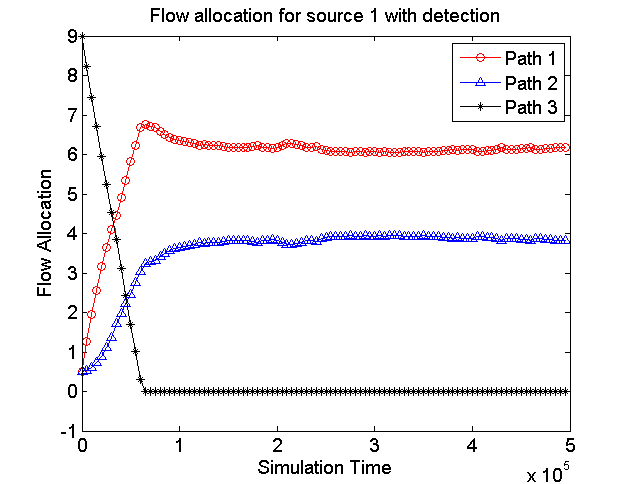} &
\includegraphics[width=2.1in]{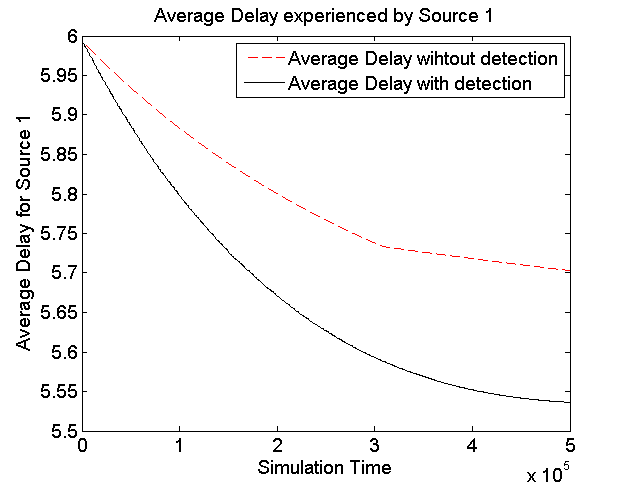}\\
\mbox{(a)} & \mbox{(b)} & \mbox{(c)} \\
\end{array}$
\caption{Simulation of our passivity framework for modeling the in-band wormhole. The time scales represent the number of iterations of the simulation. Each iteration represents a single update step of the in-band wormhole dynamics. The source rates for sources 1 and 2 are given as 10 and 5. Initial flow allocation for source 1 is [0.5,0.5,9], and flow allocation for source 2 is [0.5,0.5,4] for paths 1, 2, and 3 respectively. Link 9 is an wormhole link with falsely advertised capacity of 15. Packets allocated to path 3 will be rerouted to path 1 with probability 0.3 and to path 2 with probability 0.7. (a) The impact of an in-band wormhole on flow allocation with no mitigation mechanisms. b) The flow allocation when mitigation method is used. (c) The impact of mitigation method on average delay.} 
\label{inband}
\end{figure*}
In the in-band wormhole simulation, we assume that link 9 is an in-band wormhole. 
Upon receiving packets allocated to path 3, the malicious node allocates $\lambda$ fraction of traffic to path 1 and $1-\lambda$ fraction of traffic to path 2. This results in increased traffic to paths 1 and 2 and hence increased overall delay experienced by sources. The perceived delay for path 3 is given as
\begin{equation}
q({P_{3}}) = \lambda q(r_{P_{1}} + \lambda r_{P_{3}}) + (1-\lambda) q(r_{P_{2}} + (1-\lambda) r_{P_{3}})
\end{equation}
The mitigation mechanism is based on the anomalous delay experienced at the wormhole link. The link will be avoided if the ratio of actual delay experienced at link $l$, denoted $D_{l}$, to the expected delay exceeds a predefined threshold. The penalty of $K=10$ was added to the link price when
\begin{equation}
\log \frac{D_{l}}{f_{l}(r_{l})} > 0
\label{detection}
\end{equation}
The delay experienced at link $l$ is modeled as an exponential random variable with mean $f_{l}(r_{l})$.

We illustrate the impact of the in-band wormhole  on flow allocation (Figure \ref{inband}(a)). Packets allocated to path 3, which contains the wormhole link, are rerouted by the adversary to path 2 with probability 0.7. This results in longer delay experienced over path 2. Without mitigation, source 1 is unaware of packets flowing through wormhole tunnels, and allocates all its traffic through paths 1 and 3. Figure \ref{inband}(b) shows the flow allocation when statistical mitigation method, as described in (\ref{detection}), is used. Since the packets allocated to path 3 do not traverse a one-hop link with capacity 15 as advertised, but instead traverse a two-hop path with lower capacity, the delay will deviate significantly from its expected value.  Hence the statistical mitigation mechanism will identify the wormhole link (link 9) with high probability. This results in an equilibrium point similar to the case without wormhole. Figure \ref{inband}(c) shows the impact of mitigation on average delay. The average delay experienced by source 1 is reduced when mitigation is used. These results suggest that the sources become aware of the true topology of the network, which does not contain path 3, and achieve a Wardrop equilibrium consisting only of paths 1 and 2.

\subsection{Simulation of Impact of Out-of-band Wormhole on a Physical System }
We now study the impact of the out-of-band wormhole on a physical system. We consider a networked control system where the control loop is closed through the network shown in Figure \ref{fig:network}. The physical plant considered is a single-input, single-output integrator with dynamics given in equation (\ref{physical}). We assume the state value $x(t)$ is measured, and sampled every $h = 0.3$ time units by node $S_{1}$ and relayed to the controller node $D$. Disturbance $w(t)$ is assumed to be white Gaussian noise with zero mean and variance 1. Control gain $G=2$ is considered in the simulation.
     \begin{equation}
     \begin{array}{ll}
     \dot{x}(t) = u(t) + w(t), & t \in [kh + \tau_{k}, (k+1)h + \tau_{k+1}] \\
     u(t) = -Gx(t - \tau_{k}), & t \in [kh + \tau_{k}, (k+1)h + \tau_{k+1}] \\
     \end{array}
     \label{physical}
     \end{equation}
We consider the same M/M/1/K queue model for valid links except the propagation delay $\alpha_{l}$ for valid links is now assumed to be 0.05 time units, and propagation delay for the out-of-band wormhole link (link 9) is assumed to be 0.1 time units. Adversary controlling the wormhole link provides a low-latency link with delay 0.1 when the flow rate traversing through the wormhole is less than 5 units of flow, and once the flow rate through the wormhole link exceeds 5 units of flow, the adversary drops packets with probability 0.9.

We illustrate the impact of out-of-band wormhole on the physical system in three different cases. In the first case, we assume no mitigation strategy is employed by the network. In the second and third cases, we assume packet leash defense is employed with $\Delta_{\max} = 0.04$ and $\Delta_{\max} = 0.1$ respectively. The clock skew $\Delta$ is assumed to be an exponential random variable with mean $0.05$.

\begin{figure}[h]
	\centering
	$\begin{array}{ccc}
	\includegraphics[width=2in]{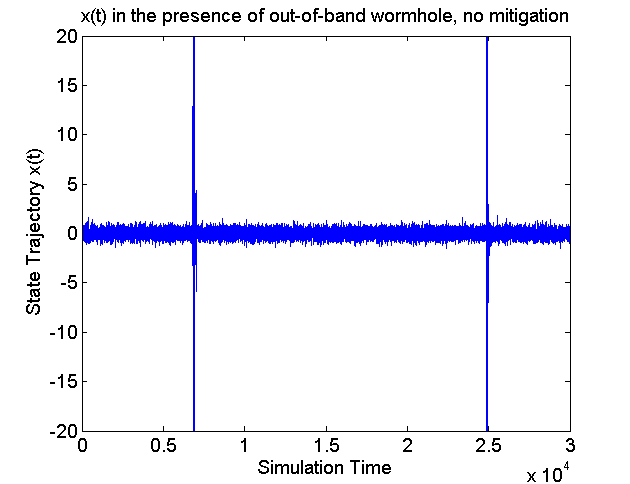} &
	\includegraphics[width=2in]{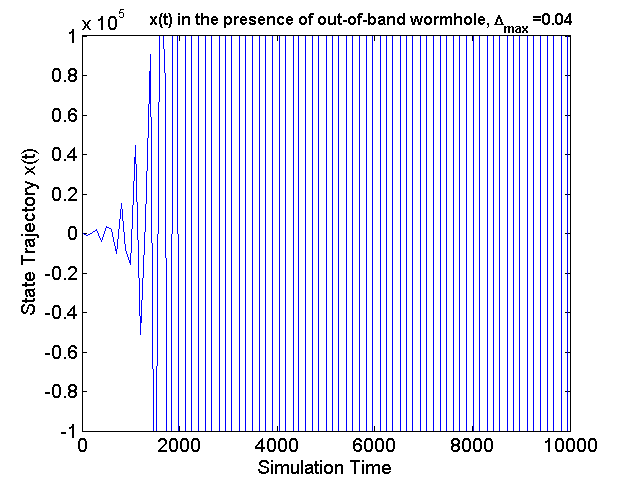} &
	\includegraphics[width=2in]{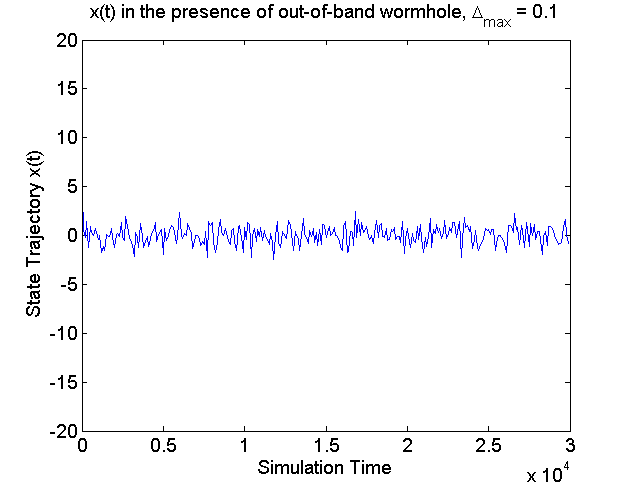} \\
	\mbox{(a)} & \mbox{(b)} & \mbox{(c)}
	\end{array}$
	\caption{State trajectory $x(t)$ in the presence of out-of-band wormhole. (a) No mitigation strategy employed (b) Packet-leash is employed with $\Delta_{\max} = 0.04$ (c) Packet-leash is employed with $\Delta_{\max} = 0.1$}
	\label{fig:simulation}
	\end{figure}

The impact of wormhole on the physical plant when no mitigation strategy is employed is illustrated in Figure \ref{fig:simulation}(a). Adversary first starts providing a low-latency link, attracting a large amount of packets traversing through the wormhole. Once the flow rate through the wormhole exceeds the threshold, the adversary drops packets with high probability, resulting in large oscillation of the plant state $x(t)$. Source node $S_{1}$ reallocates flows to paths 1 and 2 once high delay is observed on path 3, and adversary starts attracting flows again by providing low-latency link.

Figures \ref{fig:simulation}(b) and (c) illustrate the effect of mitigation strategies on the physical plant. In both cases, flows allocated to path 3 quickly converges to 0 due to packet leash. However, as shown in Figure \ref{fig:simulation}(b), a stringent packet leash with $\Delta_{\max} = 0.04$ results in growing oscillation of $x(t)$ due to overall increased delay. On the other hand,  Figure \ref{fig:simulation}(c) illustrates that when $\Delta_{\max}$ is chosen appropriately as $\Delta_{\max} = 0.1$, the plant stabilizes around the equilibrium point while successfully mitigating the wormhole attack. This case study shows that parameters of the defense mechanism need to be chosen and adjusted over time to mitigate the attack while maintaining the performance of the physical plant.

\section{Conclusion}
\label{sec:conclusion}
In this paper, we studied the wormhole attack on networked control systems, in which an adversary creates a link between two geographically distant network regions, either using a side channel, as in the out-of-band wormhole, or by colluding network nodes, as in the in-band wormhole. Using the wormhole attack, the adversary can cause violations of timing constraints in real-time systems, including dropping or delaying packets flowing into wormholes. We presented a passivity-based control-theoretic framework for modeling and mitigating the wormhole attack.  Under our framework, the flow allocation of the valid nodes, the delays experienced on the wormholes, and the wormhole mitigation algorithms were modeled as distinct, interconnected passive dynamical systems.  The passivity approach enabled us to prove stability and convergence of the system to a unique equilibrium, which satisfies the criteria for the well-known Wardrop equilibrium, under general assumptions on the adversary behavior and network mitigation mechanism. This allowed us to characterize the delays experienced by source nodes at the steady-state.

For the out-of-band wormhole attack, we quantified the increase in delay caused by the wormhole link and mapped the adversary's strategy to the optimization problem of selecting the packet-dropping rate.  We also introduced an approach for dynamically adapting the parameters of packet leash-based defenses in response to the observed network delays.  For the in-band wormhole, we used spatial statistics to estimate the delays experienced by the wormhole tunnel as a function of the number of misbehaving network nodes. In addition, we identified a new class of complex wormhole attacks consisting of both in- and out-of band wormholes, which we modeled and analyzed using our framework.

Our simulation results illustrate the trade-off between the effectiveness of the network defense and the increase in delay for the out-of-band case. In particular, we found that out-of-band wormhole causes large disturbances in the physical system by selectively dropping packets, and the parameters of packet leash defense can be chosen to reduce flow allocation to the wormhole while satisfying the delay constraint of the physical system. For the in-band case, our simulation suggests that the network defense allows the system to reach the same flow allocation equilibrium regardless of the presence of wormhole.

In our future work, we will investigate  whether the steady-state values of our passivity framework arise as equilibria of an equivalent dynamic game between the network and adversary.

\appendices
\section{Background on Passivity}
\label{sec:passive_background}
We consider a state-space model $(\Sigma)$, with state $x(t)$, input $u(t)$, and output $y(t)$, defined by
\begin{displaymath}
(\Sigma) \left\{
\begin{array}{c}
\dot{x}(t) = f(x(t), u(t)) \\
y(t) = g(x(t),u(t))
\end{array}
\right.
\end{displaymath}
The definitions and results in this subsection can be found in \cite{brogliato2007dissipative}.  A passive system is defined as follows.
\begin{definition}
\label{def:passivity}
The system $(\Sigma)$ is passive if there exists a nonnegative $C^{1}$ function $V: \mathbb{R} \rightarrow \mathbb{R}_{\geq 0}$ satisfying $V(0) = 0$ and
\begin{equation}
\label{eq:def_passivity}
\dot{V}(t) \leq -S(x(t)) + u(t)^{T}y(t)
\end{equation}
where $S(\cdot)$ is a nonnegative continuous function.  If $S(x) > 0$ for all $x \neq 0$, then the system is \emph{strictly passive}.  A function $V$ satisfying (\ref{eq:def_passivity}) for a system $(\Sigma)$ is a \emph{storage function} for $(\Sigma)$.
\end{definition}

The following two lemmas are used to construct passive systems as interconnections of passive components.
\begin{lemma}
\label{lemma:A_AT}
Suppose that the system $(\Sigma)$ is passive with $u(t) \in \mathbb{R}^{m}$ and $y(t) \in \mathbb{R}^{n}$.  Then for any $m \times n$ matrix $A$, the system $\Sigma^{\prime}$, defined by
\begin{displaymath}
(\Sigma^{\prime}) \left\{
\begin{array}{rcl}
\dot{x}(t) &=& f(x(t), A^{T}u^{\prime}(t)) \\
y^{\prime}(t) &=& Ag(x(t), A^{T}u^{\prime}(t))
\end{array}
\right.
\end{displaymath}
is passive from input $u^{\prime} \in \mathbb{R}^{n}$ to output $y^{\prime} \in \mathbb{R}^{m}$.
\end{lemma}

\begin{lemma}
\label{lemma:neg_feedback_passive}
The negative feedback interconnection of two passive systems is passive.  If at least one of the systems is strictly passive, then the negative feedback interconnection is strictly passive.
\end{lemma}

 Passivity leads to a variety of techniques for guaranteeing stability of dynamical systems, such as the following proposition.
 \begin{proposition}
 \label{prop:passive_interconnection}
 A negative feedback interconnection between two passive systems is globally asymptotically stable if at least one of the systems is strictly passive.
 \end{proposition}

 For a negative feedback interaction between two strictly passive systems with storage functions $V_{1}$ and $V_{2}$, the function $V = V_{1} + V_{2}$ is a Lyapunov function for the combined system.

\section{Proofs of Proposition \ref{prop:well_defined}, Lemma \ref{lemma:unique_Wardrop}, and Proposition \ref{prop:beta_concave}}
\label{sec:proofs}

\begin{IEEEproof}[Proof of Proposition \ref{prop:well_defined}]
We first show that $\sum_{P \in \mathcal{P}_{i}}{r_{P}(t)} = r_{i}$ for all $t > 0$.  We have that $\sum_{P \in \mathcal{P}_{i}}{r_{P}(0)} = r_{i}$ and, by (\ref{eq:Wardrop_dynamics}),
\begin{displaymath}
\frac{d}{dt}\left(\sum_{P \in \mathcal{P}_{i}}{r_{P}(t)}\right) = 0.
\end{displaymath}
Hence $\sum_{P \in \mathcal{P}_{i}}{r_{P}(t)} = r_{i}$ for all $t > 0$.  Now, suppose that for some $P \in \mathcal{P}_{i}$ and some $t > 0$, $r_{P}(t) < 0$.  Since $r_{P}$ is continuous as a function of time, $r_{P}(t^{\prime}) = 0$ for some $t^{\prime} < t$.  Define
\begin{displaymath}
t^{\ast} \triangleq \sup{\{t^{\prime} : r_{P}(t^{\prime}) = 0, \ t^{\prime} < t\}}.
\end{displaymath}
 Suppose that $P \neq P^{\ast}(t^{\ast})$.  Then
 \begin{displaymath}
 \dot{r}_{P}(t^{\ast}) = -\{q_{P}(r_{P}) - q_{P^{\ast}}(r_{P}^{\ast})\}_{+}^{r_{P}}.
 \end{displaymath}
 By definition of $P^{\ast}(t^{\ast})$, $q_{P}(r_{P}) \geq q_{P^{\ast}}(r_{P}^{\ast})$ and $r_{P}(t^{\ast}) \geq 0$, which implies that there exists $\epsilon > 0$ with $r_{P}(t^{\ast} + \epsilon) \geq 0$.  Hence, there exists $t^{\prime\prime} \in (t^{\ast}, t)$ such that $r_{P}(t^{\prime\prime}) = 0$, contradicting the definition of $t^{\ast}$.

 Now, suppose that $P = P^{\ast}(t^{\ast})$.  Then by the preceding discussion, $\dot{r}_{P^{\prime}}(t^{\ast}) \leq 0$ for all $P^{\prime} \neq P$, resulting in
 \begin{displaymath}
 \dot{r}_{P}(t^{\ast}) = -\sum_{P^{\prime} \neq P^{\ast}}{\dot{r}_{P}(t^{\ast})} \geq 0,
 \end{displaymath}
 which contradicts the definition of $t^{\ast}$.  Hence $r_{P}(t) \geq 0$ for all $P \in \mathcal{P}_{i}$ and $t \geq 0$.
 \end{IEEEproof}

 \begin{IEEEproof}[Proof of Lemma \ref{lemma:unique_Wardrop}]
 By Proposition \ref{prop:Wardrop_equilibrium}, it suffices to characterize the Wardrop equilibria of the system.  From \cite{altman2004equilibrium}, a point $\mathbf{r} = \{\mathbf{r}_{i}: i =1,\ldots,m\}$ is a Wardrop equilibrium if and only if it is a solution to the optimization problem
 \begin{equation}
 \label{eq:Wardrop_opt}
 \begin{array}{ll}
 \mbox{min}_{\mathbf{r}} & \sum_{l \in L}{h_{l}(r_{l})} \\
 \mbox{s.t.} & \sum_{P \in \mathcal{P}_{i}}{r_{P}} = r_{i}, \ i=1,\ldots,m \\
  & r_{P} \geq 0, \ P \in \cup_{i=1}^{m}{\mathcal{P}_{i}} \\
  & r_{l} = \sum_{P \ni l}{r_{P}}
 \end{array}
 \end{equation}
 where $h_{l}: \mathbb{R} \rightarrow \mathbb{R}$ is defined by
 \begin{displaymath}
 h_{l}(x) \triangleq \int_{0}^{x}{f_{l}(s) \ ds}.
 \end{displaymath}
 Since $f_{l}$ is assumed to be strictly increasing, $h_{l}$ is strictly convex.  Problem (\ref{eq:Wardrop_opt}) therefore involves minimizing a strictly convex objective function over a convex set, and hence has a unique solution.  Thus there is a unique Wardrop equilibrium, and hence there is a unique equilibrium of (\ref{eq:Wardrop_dynamics}) by Proposition \ref{prop:Wardrop_equilibrium}.
 \end{IEEEproof}

\begin{IEEEproof}[Proof of Proposition \ref{prop:neg_equivalent}]
 The dynamics (\ref{eq:Wardrop_dynamics}) with the input $q_{P}(t) = \sum_{l \in P}{f_{l}(r_{l}(t))}$ can be written as
    \begin{displaymath}
    (H) \ \left\{
    \begin{array}{ll}
   \dot{r}_{P}(t) = -\{u_{P}^{(1)} - u_{P^{min}}^{(1)}\}_{+}^{r_{P}}, & P \neq P_{i}^{min} \\
     \dot{r}_{P}(t) = -\sum_{P \neq P^{min}}{\dot{r}_{P}(t)}, & P = P_{i}^{min} \\
     y_{P}^{(1)}(t) = \dot{r}_{P}(t), & \forall P \in \mathcal{P}_{i} \\
     u_{l}^{(2)}(t) = A^{T}\mathbf{\dot{r}}(t) & \forall l \in \mathcal{L} \\
     \dot{z}_{l}(t) = u_{l}^{(2)}(t) & \forall l \in \mathcal{L}\\
     y_{l}^{(2)}(t) = f_{l}(z_{l}(t)) & \forall l \in \mathcal{L}\\
     u_{P}^{(1)}(t) = A\mathbf{y}^{(2)}(t) & \forall P \in \mathcal{P}_{i}\\
     \end{array}
     \right.
     \end{displaymath}
     Setting $\tilde{y}_{l}^{(2)}(t) = f_{l}(z_{l}(t)) - f_{l}(z_{l}^{\ast})$ and $\tilde{u}^{(1)}(t) = -A\tilde{\mathbf{y}}^{(2)}(t)$ yields the equivalent system
     \begin{displaymath}
     (\tilde{H}) \ \left\{
     \begin{array}{ll}
     \dot{\tilde{r}}_{P}(t) = -\{q^{\ast}_{P} - \tilde{u}^{1}_{P} - q^{\ast}_{P^{\ast}(q^{\ast}-\tilde{u}^{1})} + \tilde{u}^{(1)}_{P^{\ast}(q^{\ast}-u)}\}_{+}^{r_{P}}, & P \neq P^{\ast}(q^{\ast}-\tilde{u}^{1}) \\
    \dot{\tilde{r}}_{P}(t) = -\sum_{P \neq P^{\ast}(q^{\ast}-\tilde{u}^{1})}{\dot{\tilde{r}}_{P}(t)}, & P = P^{\ast}(q^{\ast}-\tilde{u}^{1}) \\
    \tilde{y}_{P}^{(1)}(t) = \dot{\tilde{r}}_{P}(t), & \forall P \in \mathcal{P}_{i} \\
    \tilde{u}^{(2)}(t) = A^{T}\mathbf{\dot{\tilde{r}}}(t) & \\
    \dot{z}_{l}(t) = \tilde{u}_{l}^{(2)}(t) & \forall l \in \mathcal{L}\\
    \tilde{y}_{l}^{(2)}(t) = f_{l}(z_{l}(t)) - f_{l}(z_{l}^{\ast}) & \forall l \in \mathcal{L}\\
    \tilde{u}_{P}^{(1)}(t) = -Ay^{(2)}(t) & \forall P \in \mathcal{P}_{i}
    \end{array}
    \right.
    \end{displaymath}
    which can be decomposed as the negative feedback interconnection of $(\tilde{H}_{1})$ and $(\tilde{H}_{2})$.
\end{IEEEproof}

 \begin{IEEEproof}[Proof of Proposition \ref{prop:beta_concave}]
First, suppose that $x^{\prime} > x$.  Define $\mathbf{E}_{x}(\cdot) = \mathbf{E}(\cdot \ | \ |\mathcal{C}| = nx)$ and $Pr_{x}(\cdot) = Pr(\cdot \ | \ |\mathcal{C}| = nx)$.  Then
\begin{IEEEeqnarray*}{rCl}
\beta(x^{\prime}) - \beta(x) &=& \mathbf{E}_{x}\left[\min{\left\{d(W_{1},W_{3}) + d(W_{3},W_{2}) : W_{3} \in \tilde{\mathcal{C}}\right\}}\right] \\
 && - \mathbf{E}_{x}\left[\min{\left\{d(W_{1},W_{3}) + d(W_{3}, W_{2}) :  W_{3} \in \tilde{\mathcal{C}}\right\}}\right].
 \end{IEEEeqnarray*}
 Since nodes are assumed to be compromised uniformly at random $Pr_{x}(\mathcal{C}) = \frac{1}{|\mathcal{C}_{x}|}$ for all $\mathcal{C} \in \mathcal{C}_{x}$, and so
 \begin{IEEEeqnarray*}{rCl}
 \beta(x^{\prime}) - \beta(x) &=& \frac{1}{|\mathcal{C}_{x^{\prime}}|}\sum_{\mathcal{C} \in \mathcal{C}_{x^{\prime}}}{\min{\{d(W_{1}, W_{3}) + d(W_{3}, W_{2}) : W_{3} \in \tilde{\mathcal{C}}\}}} \\
  && - \frac{1}{|\mathcal{C}_{x}|}\sum_{\mathcal{C} \in \mathcal{C}_{x}}{\min{\{d(W_{1}, W_{3}) + d(W_{3}, W_{2}) : W_{3} \in \tilde{\mathcal{C}}\}}} \\
 &=& \frac{1}{|\mathcal{C}_{x^{\prime}}|}\sum_{\mathcal{C} \in \mathcal{C}_{x}}{\sum_{\mathcal{C}^{\prime} \supseteq \mathcal{C}}{\min{\{d(W_{1}, W_{3}) + d(W_{3}, W_{2}) : W_{3} \in \tilde{\mathcal{C^{\prime}}}\}}}} \\
  && - \frac{1}{|\mathcal{C}_{x}|}\sum_{\mathcal{C} \in \mathcal{C}_{x}}{\min{\{d(W_{1}, W_{3}) + d(W_{3}, W_{2}) : W_{3} \in \tilde{\mathcal{C}}\}}}.
  \end{IEEEeqnarray*}
  Now, since $\mathcal{C} \subseteq \mathcal{C}^{\prime}$ in the inner summation, $$\min{\{d(W_{1}, W_{3}) + d(W_{3},W_{2}) : W_{3} \in \tilde{\mathcal{C}^{\prime}}\}} \leq \min{\{d(W_{1},W_{3}) + d(W_{3}, W_{2}): W_{3} \in \tilde{\mathcal{C}}\}},$$ which implies that
  \begin{IEEEeqnarray*}{rCl}
  \beta(x^{\prime})-\beta(x)  &\leq&  \frac{1}{|\mathcal{C}_{x^{\prime}}|}\sum_{\mathcal{C} \in \mathcal{C}_{x}}{\sum_{\mathcal{C}^{\prime} \supseteq \mathcal{C}}{\min{\{d(W_{1}, W_{3}) + d(W_{3}, W_{2}) : W_{3} \in \tilde{\mathcal{C}}\}}}} \\
  && - \frac{1}{|\mathcal{C}_{x}|}\sum_{\mathcal{C} \in \mathcal{C}_{x}}{\min{\{d(W_{1}, W_{3}) + d(W_{3}, W_{2}) : W_{3} \in \tilde{\mathcal{C}}\}}} = 0,
  \end{IEEEeqnarray*}
  as desired. To prove convexity, we have the following:
  \begin{IEEEeqnarray*}{rCl}
  \beta(x) &\triangleq& \mathbf{E}_{x}\left[\min{\{d(W_{1}, W_{3}) + d(W_{3}, W_{2}) : W_{3} \in \tilde{\mathcal{C}}\}}\right] \\
   &=& \sum_{k=0}^{\infty}{Pr_{x}(\min{\{d(W_{1},W_{3}) + d(W_{3}, W_{2}) : W_{3} \in \tilde{\mathcal{C}}\}} > k)} \\
   &=& \sum_{k=0}^{\infty}{\frac{1}{|\mathcal{C}_{x}|}\sum_{\mathcal{C} \in \mathcal{C}_{x}}{Pr\left(\bigcap_{W_{3} \in \tilde{\mathcal{C}}}{\{d(W_{1}, W_{3}) + d(W_{3}, W_{2}) > k\}}\right)}}.
   \end{IEEEeqnarray*}
   This right hand side can be rewritten as
   \begin{IEEEeqnarray*}{rCl}
    \beta(x) &=& \sum_{k=0}^{\infty}{\frac{1}{|\mathcal{C}_{x}|}\sum_{\mathcal{C} \in \mathcal{C_{x}}}{Pr\left(\bigcap_{W_{3} \in \mathcal{C}}{\left(\{d(W_{1}, W_{3}) + d(W_{2}, W_{3}) > k\} \cup \{W_{3} \in \mathcal{C}\}\right)}\right)}} \\
   &=& \sum_{k=0}^{\infty}{\frac{1}{|\mathcal{C}_{x}|}\sum_{\mathcal{C} \in \mathcal{C}_{x}}{\left(Pr(d(W_{1}, W) + d(W_{2}, W) > k, \ d(W_{1}, W_{3}) < d(W_{2},W_{3}) + 3)\right)^{nx}}},
   \end{IEEEeqnarray*}
   where we use the fact that nodes are compromised uniformly at random.  The final equality establishes that $\beta(x)$ is a nonnegative weighted sum of convex functions of $x$, and hence is convex.
\end{IEEEproof}

\bibliographystyle{IEEEtran}
\bibliography{TAC_2013}

\begin{thebibliography}{10}
\providecommand{\url}[1]{#1}
\csname url@samestyle\endcsname
\providecommand{\newblock}{\relax}
\providecommand{\bibinfo}[2]{#2}
\providecommand{\BIBentrySTDinterwordspacing}{\spaceskip=0pt\relax}
\providecommand{\BIBentryALTinterwordstretchfactor}{4}
\providecommand{\BIBentryALTinterwordspacing}{\spaceskip=\fontdimen2\font plus
\BIBentryALTinterwordstretchfactor\fontdimen3\font minus
  \fontdimen4\font\relax}
\providecommand{\BIBforeignlanguage}[2]{{%
\expandafter\ifx\csname l@#1\endcsname\relax
\typeout{** WARNING: IEEEtran.bst: No hyphenation pattern has been}%
\typeout{** loaded for the language `#1'. Using the pattern for}%
\typeout{** the default language instead.}%
\else
\language=\csname l@#1\endcsname
\fi
#2}}
\providecommand{\BIBdecl}{\relax}
\BIBdecl

\bibitem{pajic2011wireless}
M.~Pajic, S.~Sundaram, G.~Pappas, and R.~Mangharam, ``The wireless control
  network: A new approach for control over networks,'' \emph{IEEE Transactions
  on Automatic Control}, vol.~56, no.~10, pp. 2305--2318, 2011.

\bibitem{jahanian1986safety}
F.~Jahanian and A.~K.-L. Mok, ``Safety analysis of timing properties in
  real-time systems,'' \emph{IEEE Transactions on Software Engineering}, no.~9,
  pp. 890--904, 1986.

\bibitem{hu2003packet}
Y.-C. Hu, A.~Perrig, and D.~B. Johnson, ``Packet leashes: a defense against
  wormhole attacks in wireless networks,'' \emph{Twenty-Second Annual Joint
  Conference of the IEEE Computer and Communications Societies ({INFOCOM})},
  pp. 1976--1986, 2003.

\bibitem{karlof2003secure}
C.~Karlof and D.~Wagner, ``Secure routing in wireless sensor networks: Attacks
  and countermeasures,'' \emph{{Ad Hoc Networks}}, vol.~1, no.~2, pp. 293--315,
  2003.

\bibitem{kruus2006band}
P.~Kruus, D.~Sterne, R.~Gopaul, M.~Heyman, B.~Rivera, P.~Budulas, B.~Luu,
  T.~Johnson, N.~Ivanic, and G.~Lawler, ``In-band wormholes and countermeasures
  in {OLSR} networks,'' \emph{Securecomm and Workshops, 2006}, pp. 1--11, 2006.

\bibitem{poovendran2007graph}
R.~Poovendran and L.~Lazos, ``A graph theoretic framework for preventing the
  wormhole attack in wireless ad hoc networks,'' \emph{Wireless Networks},
  vol.~13, no.~1, pp. 27--59, 2007.

\bibitem{schneider2011blueprint}
F.~Schneider, ``Blueprint for a science of cyber security,'' \emph{The Next
  Wave}, vol.~19, no.~2, pp. 47--57, 2012.

\bibitem{song2011enhancement}
R.~Song, P.~C. Mason, and M.~Li, ``Enhancement of frequency-based wormhole
  attack detection,'' \emph{IEEE Military Communications Conference
  ({MILCOM})}, pp. 1139--1145, 2011.

\bibitem{mahajan2008analysis}
V.~Mahajan, M.~Natu, and A.~Sethi, ``Analysis of wormhole intrusion attacks in
  {MANET}s,'' \emph{IEEE Military Communications Conference ({MILCOM})}, pp.
  1--7, 2008.

\bibitem{baras2007intrusion}
J.~S. Baras, S.~Radosavac, G.~Theodorakopoulos, D.~Sterne, P.~Budulas, and
  R.~Gopaul, ``Intrusion detection system resiliency to byzantine attacks: The
  case study of wormholes in {OLSR},'' \emph{IEEE Military Communications
  Conference ({MILCOM})}, pp. 1--7, 2007.

\bibitem{wen2004unifying}
J.~T. Wen and M.~Arcak, ``A unifying passivity framework for network flow
  control,'' \emph{IEEE Transactions on Automatic Control}, vol.~49, no.~2, pp.
  162--174, 2004.

\bibitem{chiang2007layering}
M.~Chiang, S.~H. Low, A.~R. Calderbank, and J.~C. Doyle, ``Layering as
  optimization decomposition: A mathematical theory of network architectures,''
  \emph{Proceedings of the IEEE}, vol.~95, no.~1, pp. 255--312, 2007.

\bibitem{wang2012passivity}
Y.~Wang, V.~Gupta, and P.~J. Antsaklis, ``On passivity of networked nonlinear
  systems with packet drops,'' \emph{ISIS Technical Report}, 2012.

\bibitem{clark2012passivity}
A.~Clark, L.~Bushnell, and R.~Poovendran, ``A passivity-based framework for
  composing attacks on networked control systems,'' \emph{50th Allerton
  Conference on Communication, Control, and Computing}, 2012.

\bibitem{lee2013modeling}
P.~Lee, A.~Clark, L.~Bushnell, and R.~Poovendran, ``Modeling and designing
  network defense against control channel jamming attacks: A passivity-based
  approach,'' \emph{to appear in IEEE Annual Conference on Information Sciences
  and Systems, Workshop on Control of Cyber-Physical Systems}, March 2013.

\bibitem{altman2004equilibrium}
E.~Altman and L.~Wynter, ``Equilibrium, games, and pricing in transportation
  and telecommunication networks,'' \emph{Networks and Spatial Economics},
  vol.~4, no.~1, pp. 7--21, 2004.

\bibitem{ross2009introduction}
S.~M. Ross, \emph{{Introduction to Probability Models}}.\hskip 1em plus 0.5em
  minus 0.4em\relax {Academic Press}, 2009.

\bibitem{brogliato2007dissipative}
B.~Brogliato, O.~Egeland, R.~Lozano, and B.~Maschke, \emph{{Dissipative Systems
  Analysis and Control: Theory and Applications}}.\hskip 1em plus 0.5em minus
  0.4em\relax Springer, 2007.

\end{thebibliography}

\end{document}